\documentclass{aa}  

\usepackage{graphicx}
\usepackage{txfonts}
\usepackage[breaklinks, colorlinks, linkcolor=blue, citecolor=blue, urlcolor=blue]{hyperref}

\begin{document}

   \title{Measuring the F-corona intensity through time correlation of total and polarized visible light images}
    \titlerunning{F-corona through time correlation of total and polarized VL images}

   \author{
          A. Burtovoi\inst{1}
          \and
          G. Naletto\inst{2,3}
          \and
          S. Dolei\inst{4}
          \and
          D. Spadaro\inst{4}
          \and
          M. Romoli\inst{5,1}
          \and
          F. Landini\inst{6}
          \and
          Y. De Leo\inst{7,8}
          }

   \institute{
             INAF - Osservatorio Astrofisico di Arcetri, Largo Enrico Fermi 5, 50125, Florence, Italy\\
              \email{aleksandr.burtovoi90@gmail.com}
         \and
             Department of Physics and Astronomy, University of Padova, Via F. Marzolo 8, 35131, Padova, Italy\\
              \email{giampiero.naletto@unipd.it}
         \and
             INAF - Osservatorio Astronomico di Padova, Vicolo dell'Osservatorio 5, 35122, Padova, Italy
         \and
             INAF - Osservatorio Astrofisico di Catania, Via Santa Sofia 78, 95123 Catania, Italy
         \and
             Department of Physics and Astronomy, University of Florence, Largo Enrico Fermi 2, 50125 Firenze, Italy
         \and
             INAF - Osservatorio Astrofisico di Torino, via Osservatorio, 20, 10025 Pino Torinese, Turin, Italy
         \and
             Department of Physics and Astronomy, University of Catania, via Santa Sofia 64, 95123 Catania, Italy
         \and
             Max Planck Institute for Solar System Research, Justus-von-Liebig-Weg 3, 37077 G\"ottingen, Germany
             }

 \abstract
 { 
We present a new correlation method for deriving the F-corona intensity distribution, which is based on the analysis of the evolution of the total and polarized visible light (VL) images.
We studied the one-month variation profiles of the total and polarized brightness acquired with Large Angle Spectrometric COronagraph (LASCO-C2) and found that in some regions they are highly correlated. 
Assuming that the F-corona does not vary significantly on a timescale of one month, we estimated its intensity in the high-correlation regions and reconstructed the corresponding intensity maps both during the solar-minimum and solar-maximum periods. Systematic uncertainties were estimated by performing dedicated simulations.
We compared the resulting F-corona images with those determined using the inversion technique and found that the correlation method provides a smoother intensity distribution. 
We also obtained that the F-corona images calculated for consecutive months show no significant variation. 
Finally, we note that this method can be applied to the future high-cadence VL observations carried out with the Metis/Solar Orbiter coronagraph.
}


   \maketitle

\section{Introduction}\label{sec:1}
The brightness of the solar corona in visible light (VL) consists mainly of the contribution from the emission of 1) the K-corona, which arises from the Thomson scattering of photospheric light by free electrons, and 
2) the F-corona, which originates from diffraction or scattering of the photospheric emission by the dust grains. 
Due to the different emission mechanisms, 
these two components possess quite different properties. The K-corona is observed with numerous asymmetric bright features (e.g., streamers). It is, on average, fainter than the F-corona, especially at high altitudes ($\gtrsim$3-4 $R_\sun$, see e.g., \citealt{KoutchmyLamy1985}). Radial profiles of the K-corona emission are steeper than those of the F-corona. 
It can be shown, for example, that the intensity profiles reported in \citet{Saito1977} for the solar minimum at heliocentric distances from 2 $R_\sun$ to 5 $R_\sun$ are well described by a power-law function $f(r) \propto r^{-n}$ with the exponent $n$ equal to 
2.2 and 2.7 for the F-corona profiles along the equator and pole, respectively, and $n$ equal to 4.3 and 4.5 for those of the K-corona. 
The emission from the K-corona is highly polarized \citep{Minnaert1930}, whereas the F-corona is known to be almost unpolarized up to $\sim$20 $R_\sun$ (see discussion in \citealt{Lamy2021_C3_K_F}). According to the model of \citealt{Blackwell1966}, the polarization of the F-corona rises from 0.05\% at 5 $R_\sun$ to 0.95\% at 20 $R_\sun$. Apart from the existence of so-called dust-free zone in the close vicinity of the Sun (see \citealt{Howard2019} and references therein), the F-corona is known to be not coupled with the solar activity. \citet{Ragot2003}, however, predicted the solar cycle variations of the F-corona brightness due to the interaction between coronal mass ejections (CMEs) and the dust grains. The significance of this effect is yet to be measured. 
The F-corona is also believed to be more stable than the K-corona \citep{Morgan2007, Llebaria2021_F_SL}. Observations reveal a nearly symmetric brightness distribution of the F-corona with smooth intensity contours of super-elliptical shape, which becomes more eccentric at high elongations \citep{KoutchmyLamy1985, Stauffer2018, Llebaria2021_F_SL}. 

Both the separation of the emission from the K- and F- coronae and the accurate determination of the corresponding intensity maps are necessary to perform detailed  
studies of the electron density distribution within the solar corona and the interplanetary dust. One method for separating the K and F components is based on the inversion technique \citep{vandeHulst1950}, which assumes a certain level of symmetry in the electron density distribution. By inverting the polarized VL images, it is possible to calculate the electron density profiles and subsequently derive the image of the K-corona. The intensity map of the F-corona can be then determined as a difference between the total brightness and K-corona images (see e.g. \citealt{Saito1977, Dolei2015}). \citet{Hayes2001} proposed accounting for the contamination from the F-corona by subtracting a minimum brightness of each pixel from the total brightness images acquired over a time interval of 56 days, centered on the day of the observation. Adjusting the empirical model of the F-corona, \citet{Hayes2001} managed to minimize the contribution of the residual features of the K-corona to the F-corona map. Since this method is image-dependent, it is of limited application for the analysis  
of large data sets. More sophisticated techniques to determine  
an empirical background model of the VL images were later developed by, for example, \citealt{Morrill2006, Morgan2010, Stenborg2017} (and references therein). These techniques are appropriate for studying the dynamic structures of the K-corona such as CMEs. Another approach for separating the K- and F- coronae is used in \citet{Lamy1997, Quemerais2002, Lamy2020_C2}. In these works the intensity maps of the K-corona were determined directly from the polarized brightness images using the model of the K-corona polarization $p_K(r)$. \citet{Llebaria2021_F_SL}, in turn, performed a careful restoration of the K and F components of the solar corona using the results of \citet{Lamy2020_C2}, who carried out a detailed characterization of the polarimetric channel of the Large Angle Spectrometric COronagraph (LASCO-C2, \citealt{Brueckner1995}) -- an instrument on board the Solar and Heliospheric Observatory (SOHO, \citealt{Domingo1995}). Recently, \citet{Boe2021} presented a novel technique for separating emission from the K- and F-coronae at low altitudes (up to a few solar radii) based on the color analysis of the total solar eclipse data.

In this work, we present a new correlation method for deriving the F-corona intensity maps, which was developed using the total and polarized VL images of the LASCO instrument. We studied the evolution of the total and polarized brightness extracted from the LASCO-C2 images during a one-month time interval, and found that in some regions they are highly correlated. Assuming that the typical timescale of the F-corona variations is longer than one month, we estimated its brightness in these high-correlation regions. Our method does not require any specific assumption on the geometry of the electron density distribution and/or on the profile of the K-corona polarization. Resulting F-corona intensity maps have been compared with those obtained by means of the inversion technique. 

The paper is organized as follows. Details on the selected data set and the data reduction are reported in Sect. \ref{sec:2}. The correlation and inversion methods are described in Sect. \ref{sec:3}. The main results are summarized in Sect. \ref{sec:4}. The discussion and conclusions follow in Sects. \ref{sec:5} and \ref{sec:6}, respectively.

\section{Observations and data reduction}\label{sec:2}
In these studies, we used LASCO-C2 images acquired with the Orange filter (540-640 nm). The analyzed data set covers 
four 
months that have a high number of polarized brightness observations ($\sim$100 per month). These were May and June 2001 (solar maximum) and 
March and April 2008 (solar minimum). 

The raw data (level ``0.5'') were downloaded from the Virtual Solar Observatory (VSO)\footnote{\url{https://sdac.virtualsolar.org/cgi/search}}. The reduced and calibrated (level ``1'') total brightness images were obtained with the \texttt{SolarSoftware} package \citep{Freeland1998_SSW} routine \texttt{reduce\_level\_1.pro}. The calibrated polarized brightness images were downloaded from the corresponding LASCO data archive\footnote{\url{https://lasco-www.nrl.navy.mil/content/retrieve/polarize/}}. Each pixel of the calibrated maps provides the corona intensity in mean solar brightness (MSB) units. The images which contain large areas with zero intensity (missing blocks) were excluded from the analysis. The total number of VL images used in our analysis is reported in Table \ref{tab:obs_log}.
\begin{table}
\caption{Number of total ($B$) and polarized ($pB$) brightness images used in this work.}
\label{tab:obs_log}
\centering
\begin{tabular}{c c c c c}
\hline\hline
                & May 2001      & June 2001     & March 2008    & April 2008    \\
\hline
$B$             & 192           & 176                   & 168                   & 210             \\
$pB$    & 96            & 88                            & 85                    & 105             \\
\hline
\end{tabular}
\end{table}

In order to unify the format of total and polarized brightness images and apply the inversion technique (see Sect. \ref{sec:3.1}), we converted all maps from Cartesian ($x$, $y$) to polar ($r$, $\phi$) coordinates. 
For this, we determined the total and polarized brightness for a grid of polar coordinates, in which the radial $r$-axis ranges from 2.5 to 6.2 $R_\sun$ with a step of 0.01 $R_\sun,$ and the polar angle $\phi$-axis ranges from 0\degr~to 360\degr~with a step of 1\degr.
The polar angle is measured counterclockwise from the west solar equator.

\section{Measuring the F-corona intensity}\label{sec:3}
The total brightness ($B$) images contain the emission from the K-corona ($K$), F-corona ($F$), and scattered/stray light ($S$):
\begin{equation}
        B = K + F + S = (pK + uK) + F + S, \, 
        \label{eq:Bsum1}
\end{equation}
where $pK$ and $uK$ are the polarized and unpolarized fractions of the K-corona, respectively. 
The instrumental stray light is known to be unpolarized (see e.g., \citealt{Llebaria2021_F_SL}) and the polarization of the F-corona does not exceed 0.06\% below 6 $R_\sun$ (adopted from \citealt{Blackwell1966}). So that, even though the F-corona can be 10-100 times brighter than the K-corona at 6 $R_\sun$ (see e.g., profiles in \citealt{KoutchmyLamy1985, Dolei2015}), the polarized brightness ($pB$) maps can be considered as dominated by the emission from the K-corona, and, therefore, $pK\approx pB$. 
The systematic errors introduced by neglecting the polarization of the F-corona are discussed in Sect. \ref{sec:5.5}.

We point out that in our analysis we did not take into account the contribution from the stray light ($S=0$). As a consequence, the resulting F-corona images are contaminated by the stray light of the LASCO-C2 instrument. Separating these two components is beyond the scope of this paper. The contamination from the stray light is evaluated and discussed in Sect. \ref{sec:5.6}.

Taking into account that $pK\approx pB$ and $(F + S)\equiv F$, Eq. (\ref{eq:Bsum1}) can be then rewritten as follows:
\begin{equation}
        B = (pB + uK) + F.
        \label{eq:Bsum2}
\end{equation}

One of the possible ways to estimate $F$ is to calculate the K-corona intensity from the $pB$ map through the inversion technique (Sect. \ref{sec:3.1}) and, then, to subtract it from the total brightness image. 
Another way is to obtain $F$ directly from the data of the total and polarized brightness in the regions where $B$ and $pB$ are highly correlated (Sect. \ref{sec:3.2}).

\subsection{Inversion method}\label{sec:3.1}
\begin{figure}
        \centering
        \includegraphics[width=0.35\textwidth]{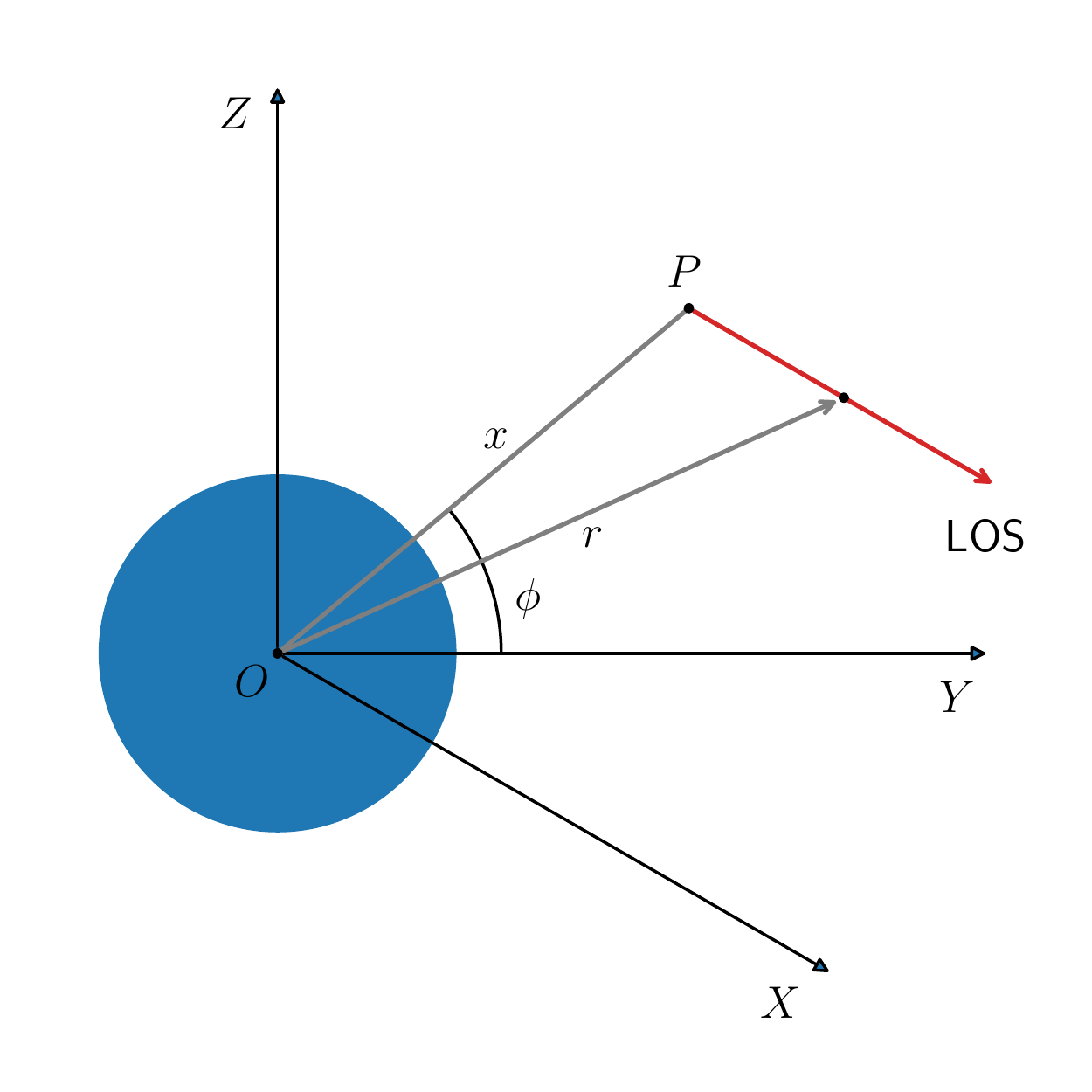} 
        \caption{Illustration of the geometry used in the inversion technique. The origin of the coordinate system ($O$) is placed at the center of the Sun. The pixel of interest $P$ is in the plane of sky $YOZ$. The line of sight (LOS) parallel to the $OX$-axis is shown as the red arrow. The $OZ$-axis points to the north pole of the Sun}
        \label{fig:inv_geom_2d}
\end{figure}

The intensity of the F-corona can be calculated using the quasi-simultaneous total and polarized brightness observations through the inversion technique (see e.g., \citealt{Saito1977, Dolei2015}). Assuming that the electron density $n_{\rm e}$ is symmetric with respect to the plane of the sky (POS) along the line of sight (LOS), and that it is a function of the heliocentric distance $r$ only, one can derive the following \citep{vandeHulst1950, Hayes2001}: 
\begin{equation}
        K(x) = C_{\rm cf} \int_x^\infty n_{\rm e}(r) \left[\left(\frac{2r^2}{x^2} -1\right)A(r) + B(r)\right] \frac{x^2}{r\sqrt{r^2-x^2}} \mathrm{d}r \,,
        \label{eq:tB}
\end{equation}
\begin{equation}
        pB(x) = C_{\rm cf} \int_x^\infty n_{\rm e}(r) [A(r) - B(r)] \frac{x^2}{r\sqrt{r^2-x^2}} \mathrm{d}r \, ,
        \label{eq:pB}
\end{equation}
where $x$ is the heliocentric distance projected on the POS and $C_{\rm cf}$ is a unit conversion factor. The adopted geometry is illustrated in Fig. \ref{fig:inv_geom_2d}. The geometric factors $A$ and $B$ can be derived from the following relations \citep{vandeHulst1950}:
\begin{equation}
        \begin{split}
                2A + B &= \frac{1-q}{1-q/3} \left [ 2(1-\cos\gamma) \right ] + \\ 
                             &+ \frac{q}{1-q/3} \left [ 1 - \frac {\cos^2\gamma}{\sin\gamma} \log \frac{1+\sin\gamma}{\cos\gamma} \right ] \,,
        \end{split}
        \label{eq:2A_plus_B}
\end{equation}
\begin{equation}
        \begin{split}
                2A - B &= \frac{1-q}{1-q/3} \left [ \frac 23 (1-\cos^3\gamma) \right ] + \\ 
                            &+ \frac{q}{1-q/3} \left [ \frac 14 + \frac{\sin^2\gamma}4 - \frac{\cos^4\gamma}{4\sin\gamma} \log \frac{1+\sin\gamma}{\cos\gamma} \right ] \, .
        \end{split}
        \label{eq:2A_minus_B}
\end{equation}
In Eqs. (\ref{eq:2A_plus_B}) and (\ref{eq:2A_minus_B}), $\sin\gamma$ corresponds to $R_\sun/r$, whereas $q$ is the coefficient of the limb darkening taken equal to 0.63 (as e.g. in \citealt{Antonucci2020_metis}).

\begin{figure}
        \centering
        \includegraphics[width=0.50\textwidth]{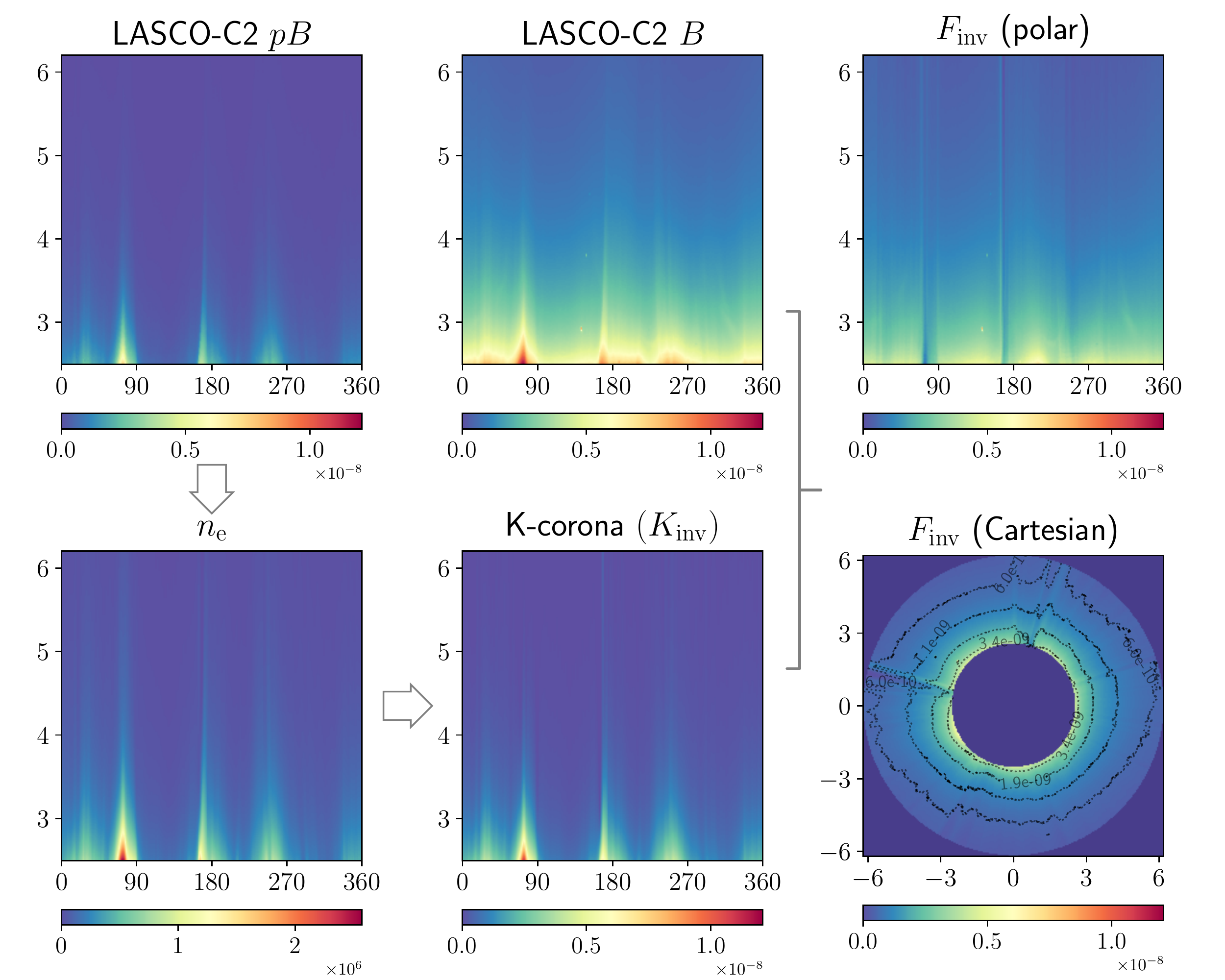} 
        \caption{Polarized ($pB$, top left panel) and total ($B$, top mid panel) brightness images obtained with LASCO-C2. $pB$ image was acquired on June 1, 2001 at 9:00, whereas the $B$ image was acquired on June 1, 2001 at 09:07. The electron density $n_{\rm e}$ (bottom left panel), K-corona $K_{\rm inv}$ (bottom mid panel) and F-corona $F_{\rm inv}$ maps (right panels) are calculated using the inversion technique (see Sect. \ref{sec:3.1}). $F_{\rm inv}$ is shown in polar and Cartesian coordinates. The horizontal axis corresponds to the polar angle $\phi$ in degrees, whereas the vertical axis is the heliocentric distance $r$ in the $R_\sun$ units in all panels except the bottom right, for which both axes correspond to the heliocentric distance. The color bar represents the intensity in MSB units in all panels except the $n_{\rm e}$ map, for which it represents the density in cm$^{-3}$ units. 
        }
        \label{fig:inv}
\end{figure}

We calculated the electron density distribution inverting the polarized brightness images as explained in \citet{Hayes2001} and \citet{Dolei2015}. For this, we determined the radial profiles $n_{\rm e}(r)$ of the electron density assuming that they follow the polynomial form: $n_{\rm e} = \sum_{k=1}^4 \alpha_kr^{-k}$. We substituted $n_{\rm e}$ in Eq. (\ref{eq:pB}) with the polynomial expression and performed the least-squares fitting of the resulting function $pB(x)$ to the polarized brightness profile observed with LASCO-C2 at each polar angle $\phi$. As a result, we obtained the coefficients $\alpha_k$ and, therefore, the density profiles $n_{\rm e}(r)$. For further details on this procedure, we refer the reader to \citet{Hayes2001}. Then, we determined the K-corona intensity map $K_{\rm inv}$ by means of Eq. (\ref{eq:tB}). The F-corona model ($F_{\rm inv}$) was calculated as the difference between the total brightness image that follows the $pB$ acquisition of interest and the K-corona map. In the typical LASCO-C2 acquisition sequences the time separation between polarized and the immediately following total brightness images is on the order of $\sim$10 minutes. An example of resulting images of $n_{\rm e}$, K-corona, and F-corona, together with the LASCO-C2 $B$ and $pB$ intensity maps are shown in Fig. \ref{fig:inv}.

F-corona maps calculated with this technique usually have distinct faint features (e.g., at $\phi$ equal to $\sim$90\degr~and $\sim$180\degr~in Fig. \ref{fig:inv}). 
However, arising from the scattering of photospheric light by the interplanetary dust, 
the F-corona  is expected to show a smoother and more symmetric brightness distribution without prominent features appeared at certain polar angles. 
It is, therefore, not likely that these features are intrinsic to the F-corona.
Being positionally coincident with the bright streamers visible in the $pB$ and K-corona maps, they are most probably spurious and correspond to the residuals of the K-corona. As discussed in \citet{Dolei2015}, these features appeared due to the oversubtraction of the K-corona in the specific regions. In fact, in order to apply the inversion technique, we assumed that for each polar angle $\phi$ there is a certain relation between electron density $n_{\rm e}$ and heliocentric distance $r$, which implies a symmetry in the electron distribution with respect to the POS. 
Such an approximation is definitely not appropriate for describing asymmetrical structures such as streamers. 
Using it for calculating the intensity of the K-corona can lead to its overestimation (or in some cases to the underestimation) and, therefore, to a wrong reconstruction of the F-corona.

\subsection{Correlation method}\label{sec:3.2}
\begin{figure}
        \centering
        \includegraphics[width=0.50\textwidth]{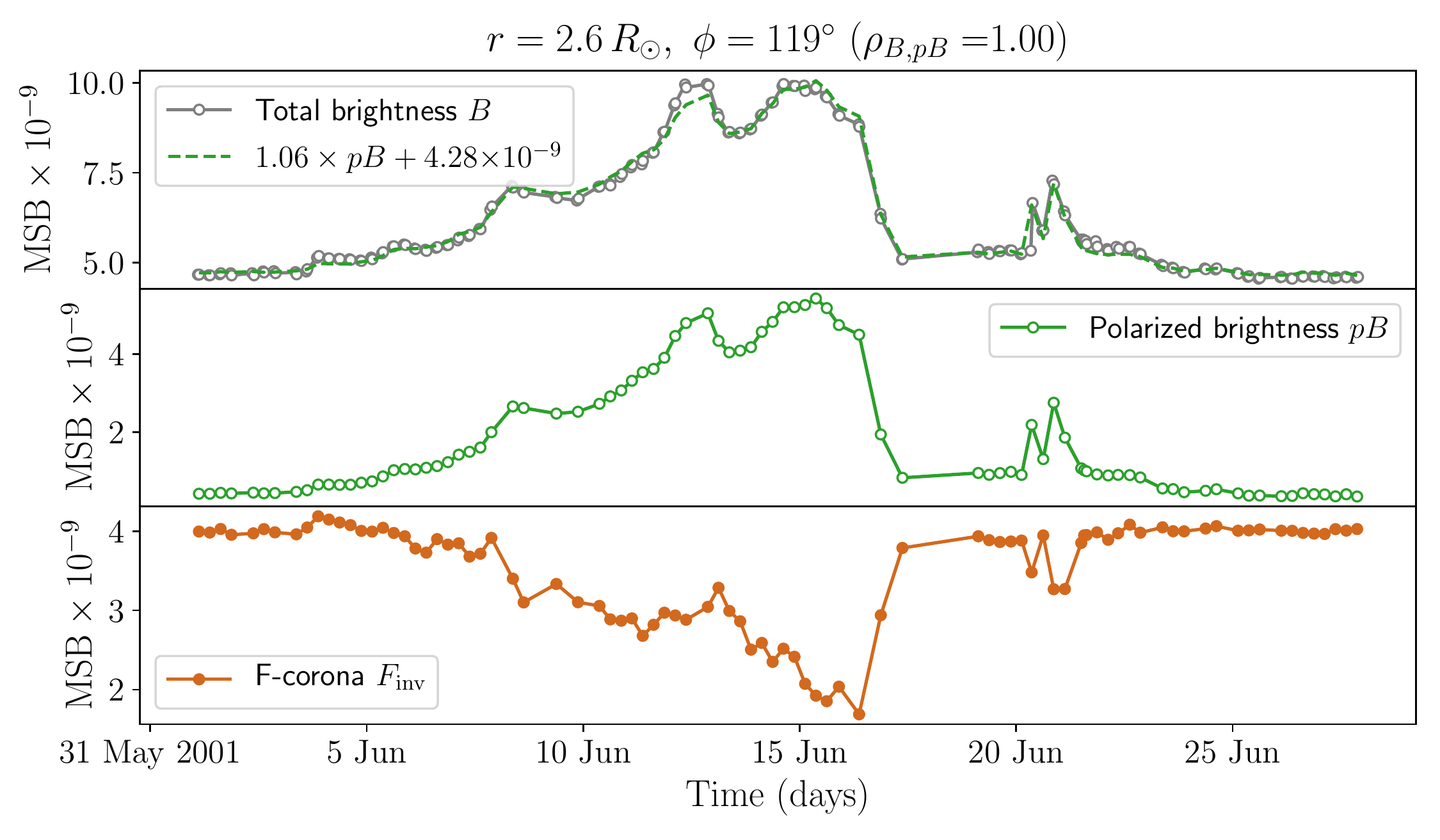} 
        \caption{Evolution of intensity of the total brightness $B$ (top panel), polarized brightness $pB$ (mid panel), and F-corona $F_{\rm inv}$ (bottom panel) measured in the pixel ($r=2.6$ $R_\sun$, $\phi=119\degr$) in a sequence of LASCO-C2 images. The dashed green line in the top panel shows the $pB$ profile re-scaled according to the linear model best fitting the $B$-$pB$ regression (see Fig. \ref{fig:pix_corr}). The third panel shows the intensity of the F-corona $F_{\rm inv}$ calculated using the inversion method (see Sect. \ref{sec:3.1} for details). The $x$-axis is in units of days and covers one month (June 2001), whereas the $y$-axis is in MSB units. The Pearson's correlation coefficient of the total and polarized brightness ($\rho_{B,pB}$) is equal to 1.}
        \label{fig:pix_evo}
\end{figure}

\begin{figure}
        \centering
        \includegraphics[width=0.4\textwidth]{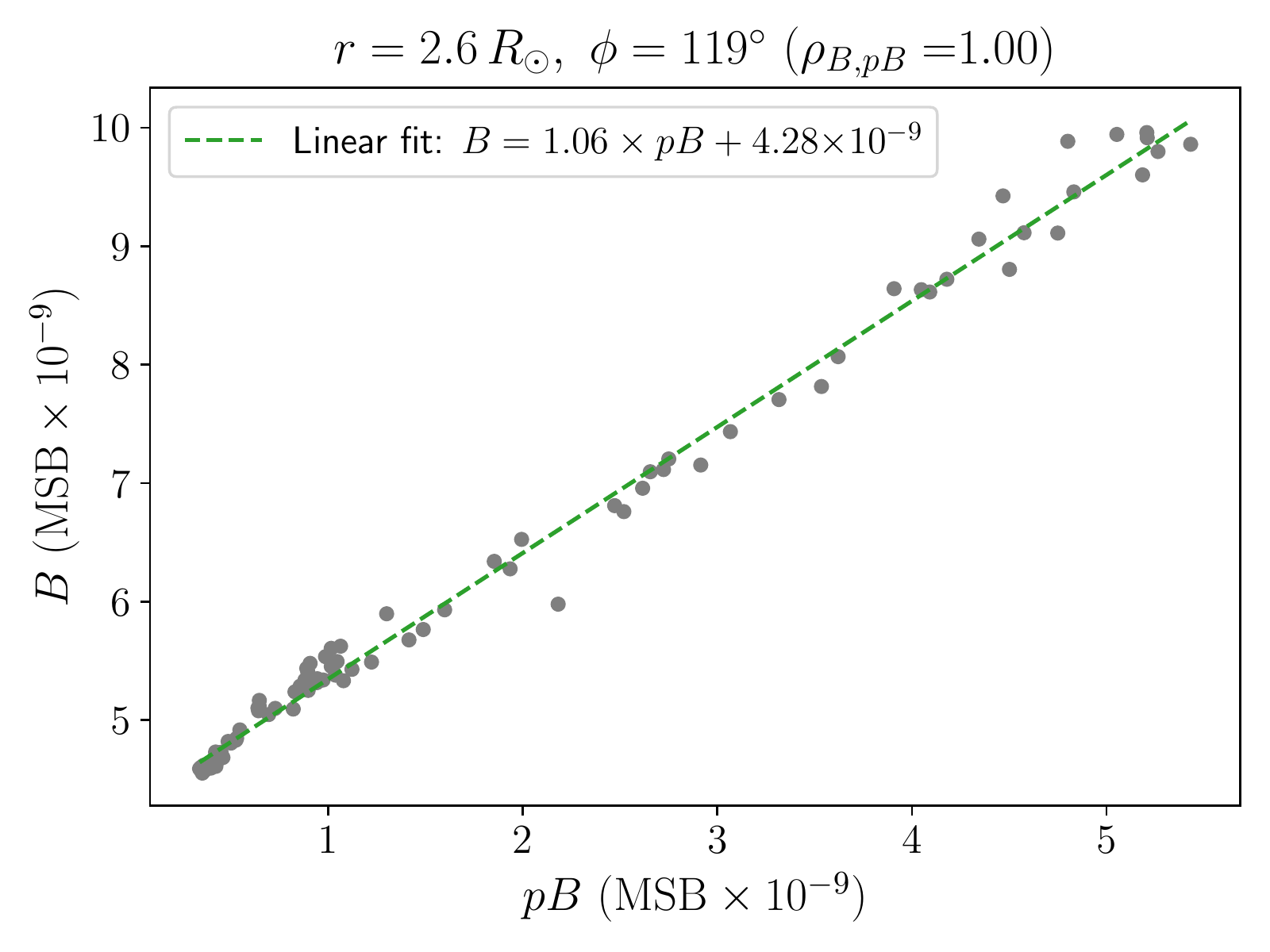}  
        \caption{Regression between intensity of total ($B$) and polarized brightness ($pB$) obtained in June 2001 at $r=2.6$ $R_\sun$ and $\phi=119\degr$ (see also Fig. \ref{fig:pix_evo}). $x$ and $y$ axes are in MSB units. The dashed green line shows the best fitting linear function.}
        \label{fig:pix_corr}
\end{figure}

\begin{figure}
        \centering
        \includegraphics[width=0.45\textwidth]{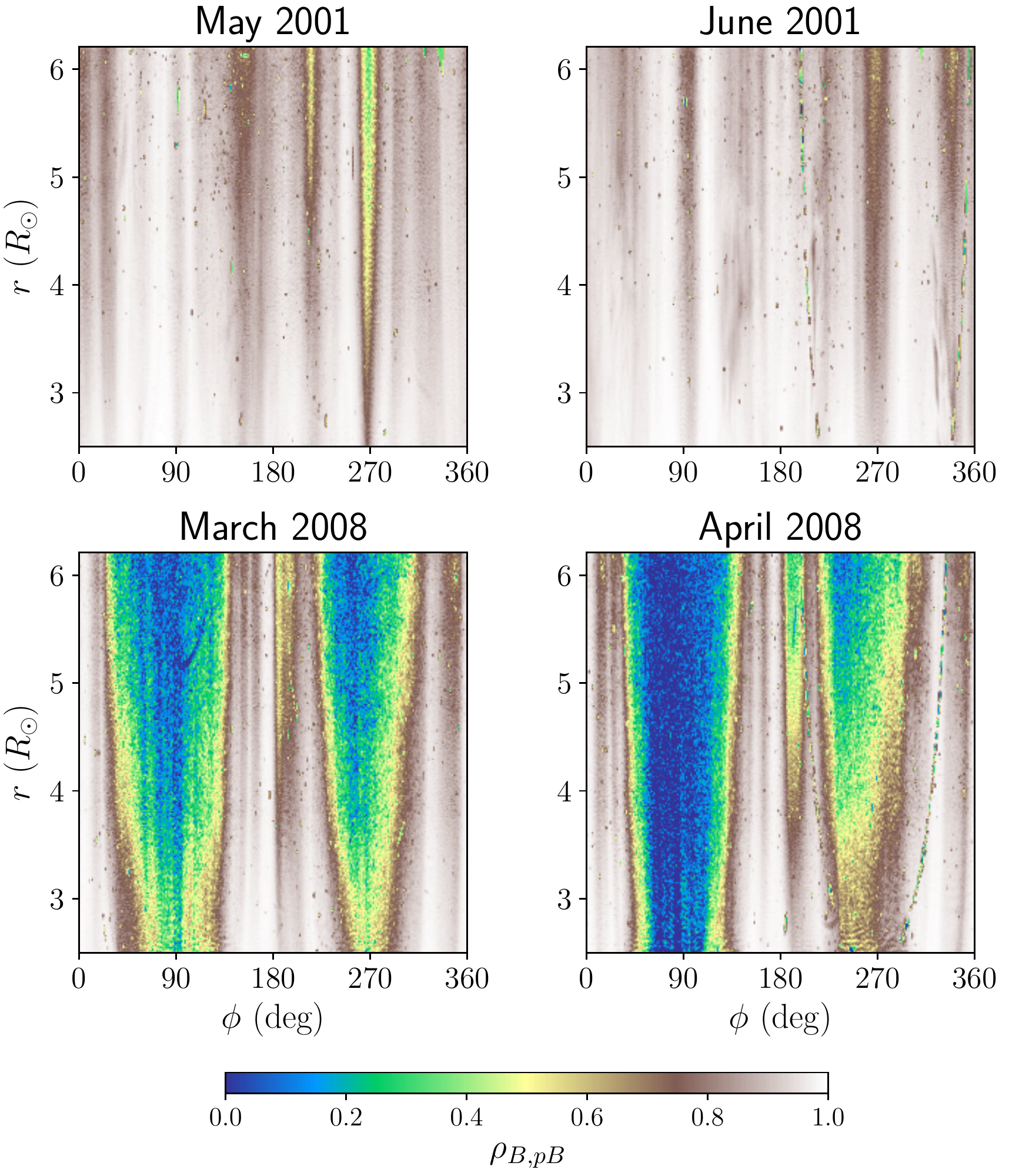} 
        \caption{Correlation coefficient maps obtained for May and June 2001 (solar maximum, top panels) and for 
March and April 2008 (solar minimum, bottom panels). 
In all panels the horizontal axis corresponds to the polar angle $\phi$ in degrees, and the vertical axis is the heliocentric distance $r$ in the $R_\sun$ units. The colors represent the value of the Pearson's correlation coefficient $\rho_{B,pB}$ which ranges between 0 and 1. 
}
        \label{fig:corr_maps}
\end{figure}

Our method for the reconstruction of the F-corona is based on the analysis of the evolution of the total and polarized brightness images.
For each pixel of the $B$ and $pB$ intensity maps, we compared the temporal behavior of the total and polarized brightness, for example over one month, and calculated the Pearson's correlation coefficient $\rho_{B,pB}$ (see, as an example, Figs. \ref{fig:pix_evo} and \ref{fig:pix_corr}). The value of the total brightness at the time of the polarized brightness acquisition is calculated by performing a linear interpolation of the preceding and following measurements of $B$. As shown in Fig. \ref{fig:corr_maps}, the regions where $B$ and $pB$ are highly correlated ($\rho_{B,pB}$ close to 1) cover 
almost the full LASCO-C2 FOV during the solar maximum period and about a half of it during the solar minimum period. 
For each high-correlation pixel, the relation between the total and polarized brightness can be reasonably well approximated with a linear function (see Fig. \ref{fig:pix_corr}):
\begin{equation}
        B = a \times pB + b = f_B (pB)\,,
        \label{eq:linfun}
\end{equation}
where $a$ and $b$ are constants for a given pixel and considered time interval, which are calculated fitting the corresponding $B$-$pB$ regression.

On the other hand, the total brightness is equal to the sum of the intensity of the K- and F- coronae:
\begin{equation}
        B(t) = K(t) + F\,,
        \label{eq:Beq1}
\end{equation}
where we highlighted the time dependence of $B$ and $K$. We assumed that the typical time scale of the F-corona variations is longer than one month, and, therefore, $F$ can be considered as a constant in our analysis (see e.g., \citealt{Morgan2007, Llebaria2021_F_SL}). 

We note that the F-corona brightness seen with LASCO 
evolves with time due to the movement of SOHO on the elliptical orbit, which is also inclined with respect to the plane of symmetry of the inner zodiacal cloud (see detailed discussion in \citealt{Llebaria2021_F_SL}). In addition, rotating around the Sun, LASCO observes the F-corona emission originated from different parts of the dust cloud. Using restored F-corona maps provided by \citet{Llebaria2021_F_SL}, we checked that on a one-month timescale such effects introduce intensity variations which are less than $\sim$1\%.

Considering that the polarized brightness $pB$ is a fraction of the K-corona ($K=\alpha\,pB$, $\alpha\ge1$), Eq. (\ref{eq:Beq1}) can be rewritten as follows: 
\begin{equation}
        B(t) = \alpha(t) \times pB(t) + F.
        \label{eq:Beq2}
\end{equation}
Since $B$ and $pB$ are highly correlated and show a linear relation between each other, we can change the argument in Eq. (\ref{eq:Beq2}) from time $t$ to intensity $pB$:
\begin{equation}
        B(pB) = \alpha(pB) \times pB + F.
        \label{eq:Beq3}
\end{equation}

Since the left hand sides of Eqs. (\ref{eq:linfun}) and (\ref{eq:Beq3}) are the same, we can write the following differential equation for $\alpha$:
\begin{equation}
        \mathrm{d}\alpha/(a-\alpha) = \mathrm{d}pB/pB.
        \label{eq:dalpha}
\end{equation}
Solving it, we obtain the relation between $\alpha$ and $pB:$ 
\begin{equation}
        \alpha = a + C/pB \,,
        \label{eq:dalpha_sol}
\end{equation}
where $C$ is the integration constant. Since $pB$ evolves with time, Eq. (\ref{eq:dalpha_sol}) represents the time dependence of the parameter $\alpha$. Additional details and an interpretation of the relation between $\alpha$, $a,$ and $C$ are provided in Appendix \ref{sec:app_alpha}.

In the particular case when $C=0$, $\alpha$ does not depend on time and is equal to $a$. As follows from Eqs. (\ref{eq:tB}) and (\ref{eq:pB}), this situation is possible when the electron density is changing with time while its spatial distribution remains the same: $n_{\rm e}= N(t)\,f(r)$. Since the solar corona is constantly rotating, strictly speaking, such a situation can be excluded from our analysis.

In all other cases, $C\ne0$, and the linear relation between $B$ and $pB$ (Eq. (\ref{eq:Beq3})) can be rewritten as
\begin{equation}
        B(pB) = a \times pB + C +  F.
        \label{eq:Beq4}
\end{equation}
Comparing Eqs. (\ref{eq:linfun}) and (\ref{eq:Beq4}), we can conclude that the parameter $b$ obtained from the linear fit is equal to $b=C +  F$. Using only the total and polarized brightness images, it is not possible to disentangle the value of $C$ from $F$. In order to constrain $C$ and subsequently estimate the intensity of the F-corona, we performed a number of simulations of rotating streamers.

\subsubsection{Constraining $C$ by the simulations of the rotating streamer(s)}\label{sec:3.2.1}
We performed simulations of non-evolving streamers of conic shape, which rotate solidly with the solar corona in a vacuum environment. 
The electron density distribution within each streamer was approximated with a power-law function of the cone height  $n_{\rm str} \propto h^{-5}$ (adopted from \citealt{Gibson1999, Antonucci2005}). It decreases along the height toward the vertex of the cone.
The simulations cover 25 days (almost full rotation) with a cadence similar to that of the LASCO-C2 $pB$ observations (three times per day).

\begin{figure}
        \centering
        \includegraphics[width=0.5\textwidth]{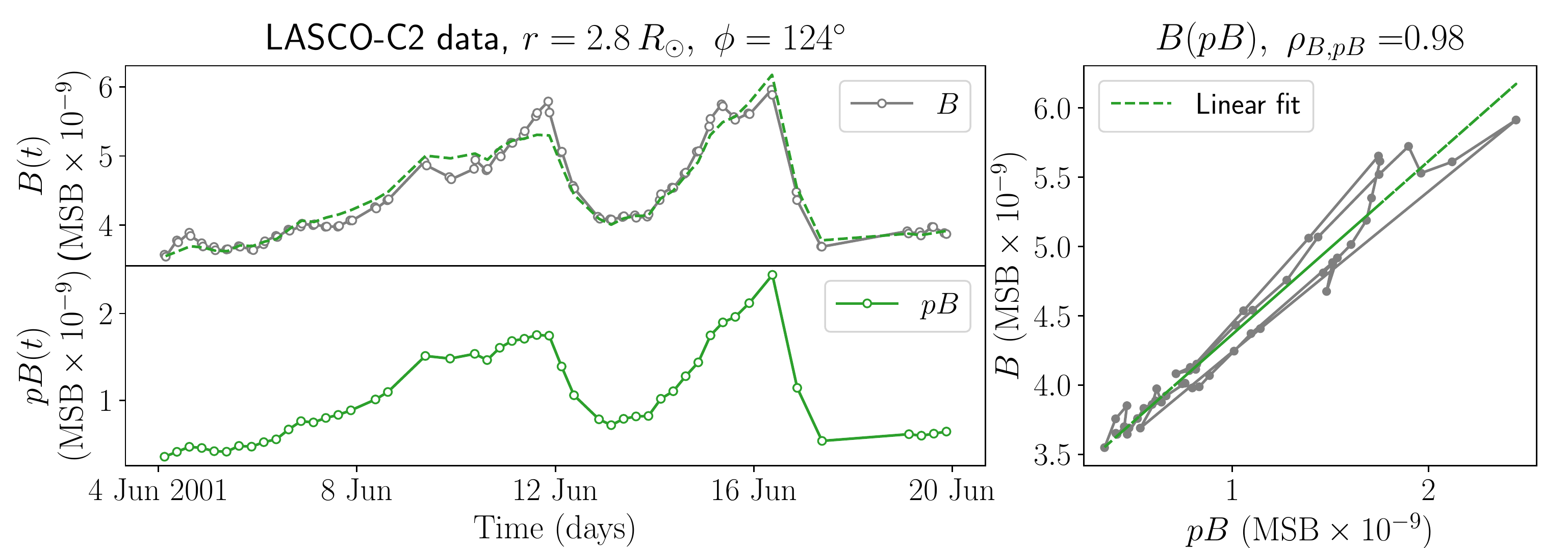} 
        \includegraphics[width=0.5\textwidth]{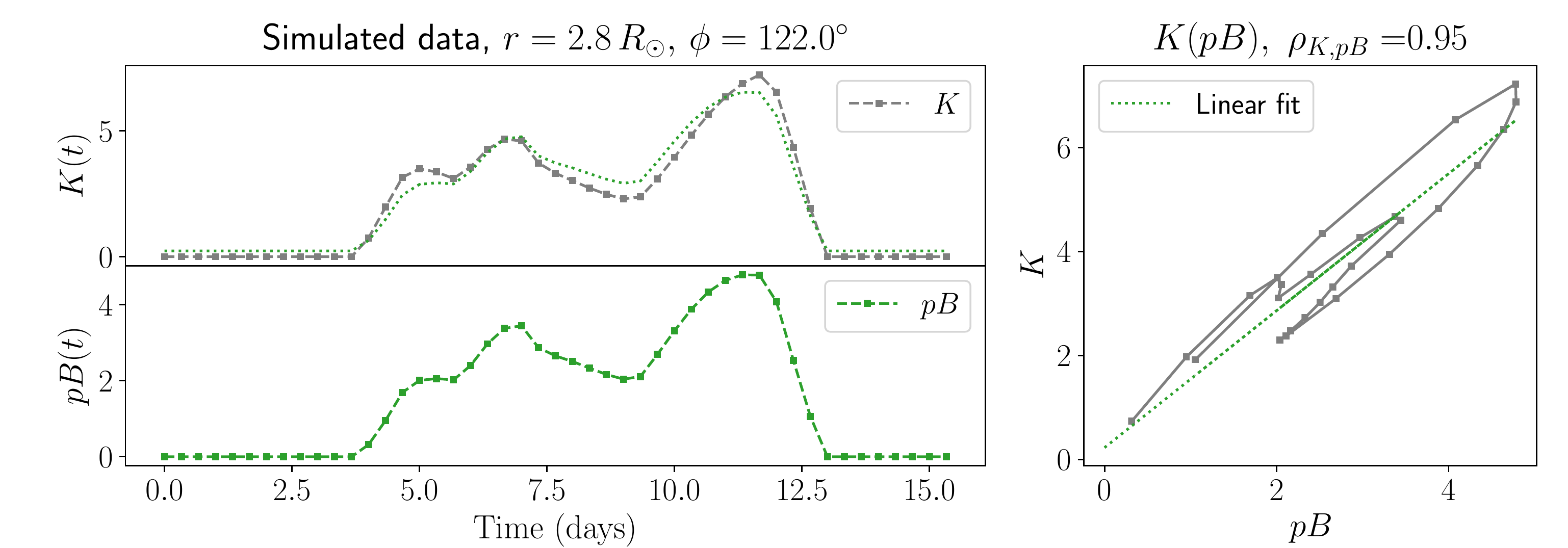} 
        \includegraphics[width=0.43\textwidth]{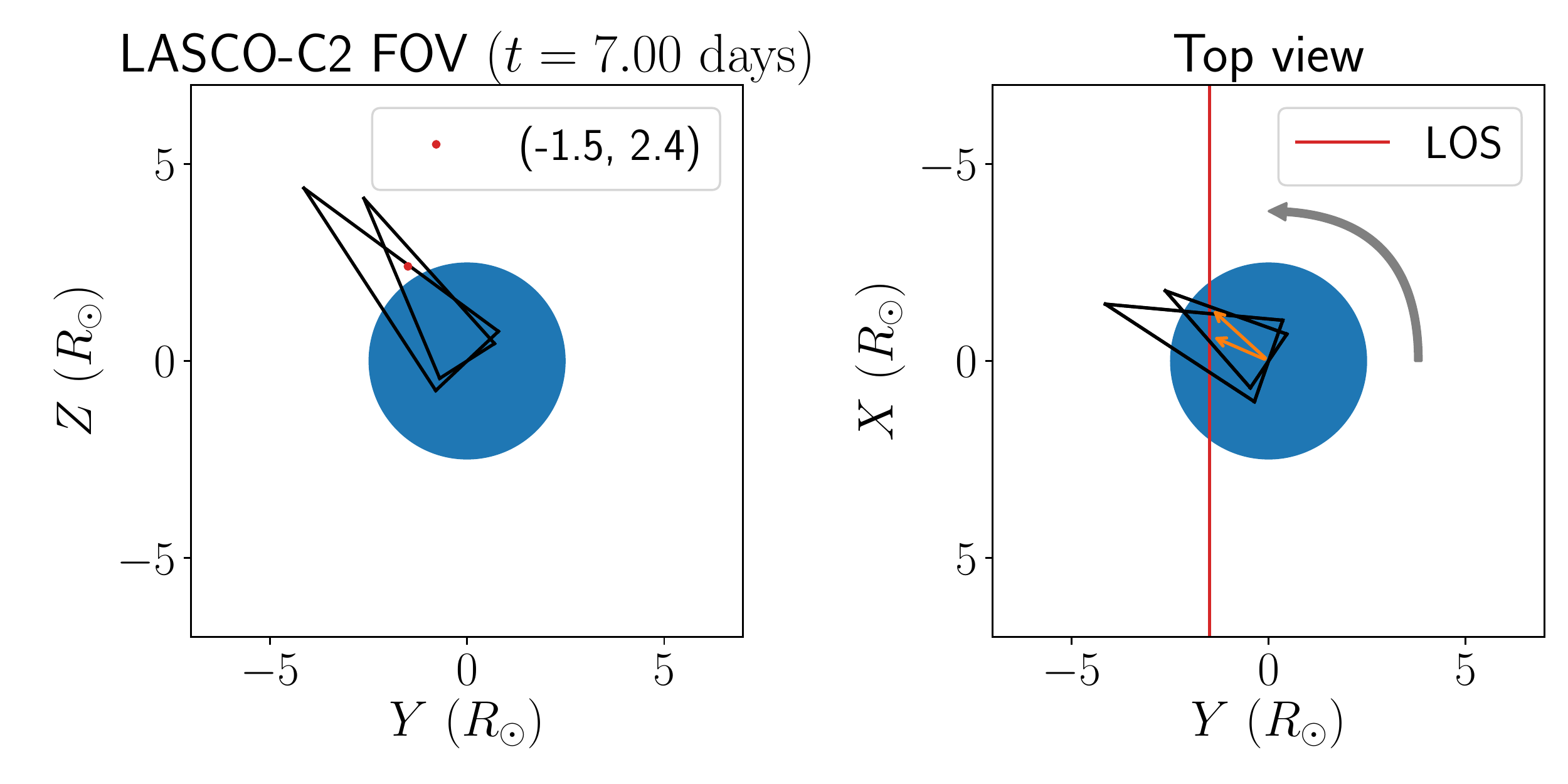} 
        \caption{Comparison of variation profiles obtained with observations and simulations. Top panels: Variation profiles of $B$ and $pB$ from the LASCO-C2 data (June 2001) and their correlation measured in the pixel at $r=2.8$ $R_\sun$ and $\phi=124\degr$. Mid panels: Variation profiles of $K$ and $pB$ and their correlation obtained by the simulation ``Sim1'' (see Table \ref{tab:all_sims}) at the similar location. 
        The dashed and dotted green lines correspond to the $pB$ profile rescaled according to the linear model best fitting the $B$-$pB$ and $K$-$pB$ regressions, respectively. 
        The real and simulated observations cover the time interval of $\sim$15 days. The simulated intensities are reported in the arbitrary units. Bottom panels represent the orientation of the two broad simulated streamers at $t=7$ days. The position of the pixel of interest is marked with the red dot. $X$-, $Y$-, and $Z$- axes are in the $R_\sun$ units. The gray arrow shows the direction of the solar rotation. The orange arrows in the bottom right panel are the radius vectors $\mathbf{r}_1$ and $\mathbf{r}_2$ from the center of the Sun to the point where the LOS enters and exits a streamer, respectively (see text for details).}
        \label{fig:sim_val_int1_broad}
\end{figure}

For each given pixel of the LASCO-C2 FOV at the projected heliocentric distance $x$, we determined the radius vectors $\mathbf{r}_1$ and $\mathbf{r}_2$ from the center of the Sun to the point where the LOS enters and exits a streamer, respectively (see arrows in the bottom right panel of Fig. \ref{fig:sim_val_int1_broad}). 
Projecting vector ($\mathbf{r}_1-\mathbf{r}_2$) to the height of a streamer, we calculated the relation between $n_{\rm str}$ and heliocentric distance $r$.
Using the $r_1$ and $r_2$ values as the integration limits in Eqs. (\ref{eq:tB}) and (\ref{eq:pB}), we calculated the $K_{\rm str}$ and $pB_{\rm str}$ intensities:
\begin{equation}
        K_{\rm str}(x) = \frac12 \int_{r_1}^{r_2} n_{\rm str} \left[\left(\frac{2r^2}{x^2} -1\right)A(r) + B(r)\right] \frac{x^2}{r\sqrt{r^2-x^2}} \mathrm{d}r \,,
        \label{eq:K_sim}
\end{equation}
\begin{equation}
        pB_{\rm str}(x) = \frac12 \int_{r_1}^{r_2} n_{\rm str} \, [A(r) - B(r)] \frac{x^2}{r\sqrt{r^2-x^2}} \mathrm{d}r \, .
        \label{eq:pB_sim}
\end{equation}
For simplicity, we omitted the conversion factor $C_{\rm cf}$. We also added a factor $(1/2)$ in both equations, since in our simulations the electron density distribution is not symmetric with respect to the POS. We note that if the vector ($\mathbf{r}_1-\mathbf{r}_2$) intersects the POS, the integrals in Eqs. (\ref{eq:K_sim}) and (\ref{eq:pB_sim}) are split in two: the first one with the integration limits ($x$, $r_1$) and the second one with ($x$, $r_2$). Performing calculations for each moment of time $t$ and for all streamers inside the FOV, we obtained the evolution of $K(t)$ and $pB(t)$. 

In order to validate this approach, we compared the evolution of total and polarized brightness extracted from the LASCO-C2 data with $K(t)$ and $pB(t)$ simulated using different number of streamers of various sizes and orientations and considering different pixels within the LASCO-C2 FOV (see Table \ref{tab:all_sims}). Aiming to reproduce the main features seen in the data, we simulated large streamers of $\sim$6 $R_\sun$ height and $\sim$10\degr--20\degr~angular aperture at the vertex. For the particular case of the polar streamer, we simulated a narrow cone with an aperture of 2.5\degr~(see Fig. \ref{fig:sim_val_pol_narrow} and Table \ref{tab:all_sims}). We found that the simulations can reproduce the shape of the variation profiles reasonably well, together with the average slope of the corresponding regression and its correlation coefficient (see some examples in Fig. \ref{fig:sim_val_int1_broad} and Appendix \ref{sec:app_sim_tab_fig}).

In order to minimize the effect of the solar rotation, we divided the simulated intensity profiles into the overlapping three-day segments and analyzed each segment separately (see Sect. \ref{sec:3.2.2} for details). For segments with the correlation coefficient $\rho_{K,pB}$ close to 1 ($\rho_{K,pB}\ge0.85$), we fitted the $K$-$pB$ regression with a linear function $f_K (pB)$. Since in our simulations we did not include the contribution from the F-corona, the constant term of $f_K (pB)$ corresponds to the parameter $C$:
\begin{equation}
        f_K (pB) = a \times pB + C\,.
        \label{eq:linfun_sim}
\end{equation}
We excluded the segments with an average slope of the $K$-$pB$ regression smaller than 0.95 ($a<0.95$) from the analysis. In fact, as follows from Eq. (\ref{eq:dalpha_sol}), if $a$ is less than 1 the parameter $|C|$ will reach a large value in order to satisfy the condition $\alpha\ge1$. 
In our procedure, we actually decreased the threshold from 1 to 0.95, aiming to include the segments with low $|C|$, 
for which the best fitting parameter $a$ appeared to be close to but smaller than 1 due to the presence of some fluctuations.

\begin{figure}
        \centering
        \includegraphics[width=0.5\textwidth]{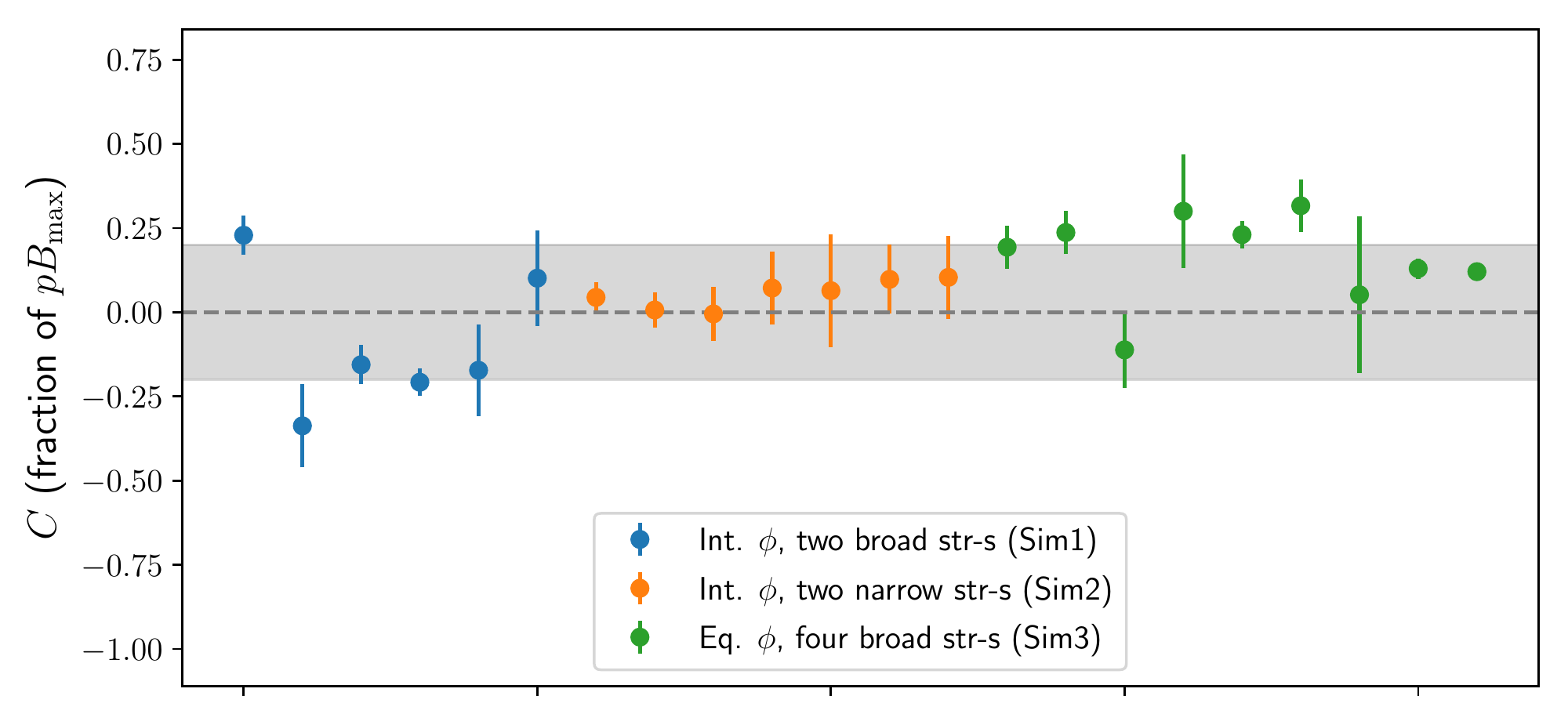} 
        \caption{Normalized values of the parameter $C$ obtained by fitting the simulated $K$-$pB$ regression. Blue, orange, and green colors mark the three-day segments of multi-streamer simulations at intermediate (Int.) and equatorial (Eq.) polar angles, as shown in Figs. \ref{fig:sim_val_int1_broad}, \ref{fig:sim_val_int2_narrow}, and \ref{fig:sim_val_eq_broad}, respectively. The simulations are labeled as in Table \ref{tab:all_sims}. The error bars represent the best-fitting 1-$\sigma$ uncertainties. The $y$-axis shows the values of $C$ as a fraction of $pB_{\rm max}$ in a segment. The gray shaded area highlights the range of $\pm20\%$ of $pB_{\rm max}$.}
        \label{fig:Csim_combined_strs_GOOD} 
\end{figure}

\begin{figure}
        \centering
        \includegraphics[width=0.5\textwidth]{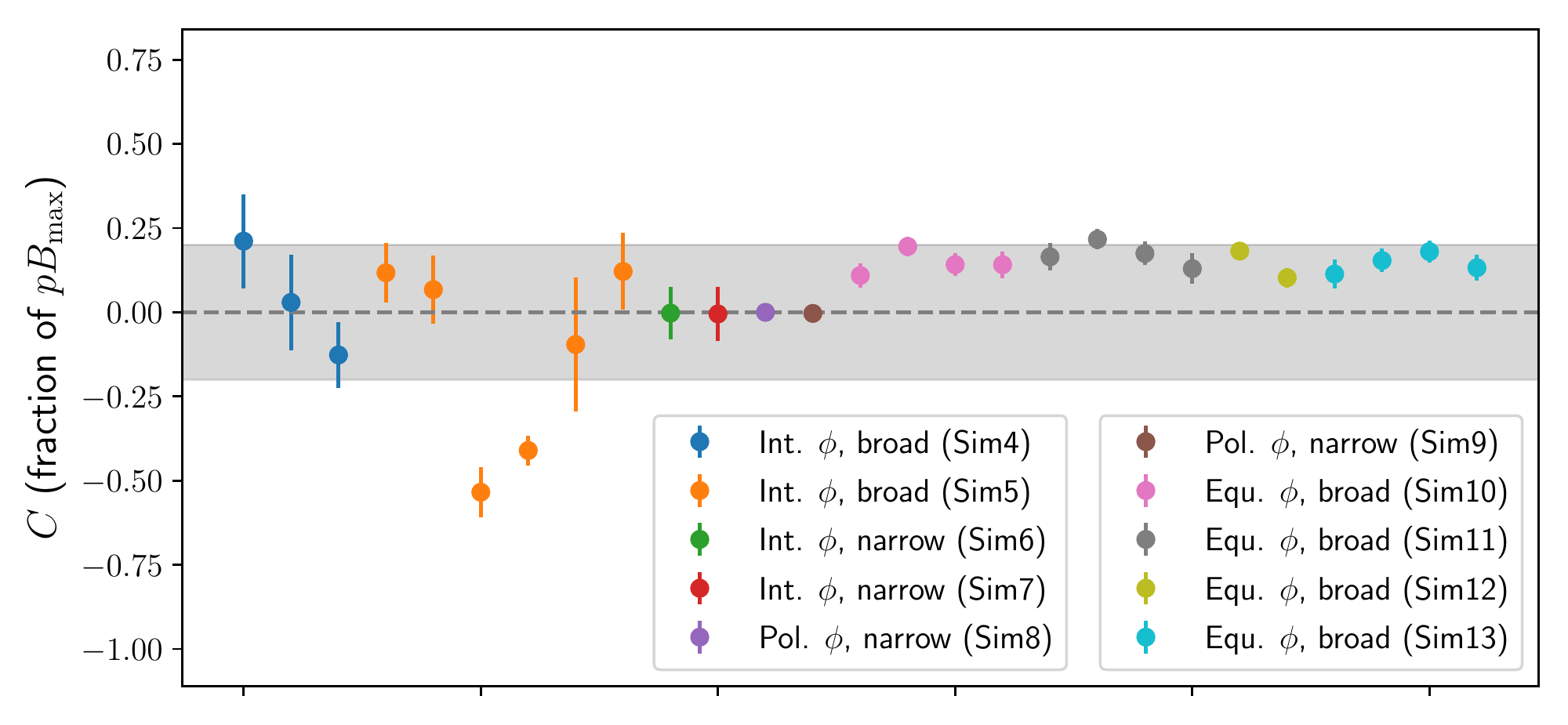} 
        \caption{Same as Fig. \ref{fig:Csim_combined_strs_GOOD}, but for different single-streamer simulations at intermediate (Int.), polar (Pol.), and equatorial (Equ.) polar angles. Each color marks the three-day segments of a specific simulation labeled as in Table \ref{tab:all_sims}.}
         \label{fig:Csim_single_strs_GOOD}
\end{figure}

Since the simulated intensities are calculated in arbitrary units, we normalized the parameter $C$ to the maximum $pB$ value reached in a segment ($pB_{\rm max}$). We then compared the normalized values of $C$ calculated for different segments and simulations and found that in many cases it stays within the range of $\pm20\%$ of $pB_{\rm max}$ (see Figs. \ref{fig:Csim_combined_strs_GOOD} and \ref{fig:Csim_single_strs_GOOD}). 

Although in the rotating corona the geometry of the electron distribution is constantly changing, 
in many cases $C$ is close to 0. For example, 
the simulations of non-equatorial and narrow streamers (with an aperture of $\lesssim$10\degr) provide the best-fitting value of $C\approx0$ and its 1-$\sigma$ uncertainty of $\lesssim20\%$ of $pB_{\rm max}$. 
Such streamers intersect the LOS in a short time interval, so the variation of $\alpha$ is not large, and, therefore, $C$ is close to 0. The smallest values of $|C|$ are obtained in the polar regions, where the effect of the solar rotation is less significant and the polarized fraction of the K-corona remains rather constant in time (see e.g., purple and brown points in Fig. \ref{fig:Csim_single_strs_GOOD}). 
On the other hand, by simulating equatorial and/or broad streamers, we found that some segments can provide the value of $C$ outside the range of $\pm20\%$ of $pB_{\rm max,}$ even when considering the uncertainties.

\subsubsection{Estimating the F-corona intensity with the LASCO-C2 data}\label{sec:3.2.2}
Similarly to the approach described in Sect. \ref{sec:3.2.1}, we divided one-month intensity profiles extracted from the LASCO-C2 data into three-day segments (equivalent to $\sim$40\degr~of the solar rotation) and analyzed each of them separately. Given a cadence of the $pB$ acquisitions during the chosen months of the LASCO-C2 data set ($\sim$3 times per day), a three-day time interval is sufficiently long for accumulating several measurements and performing an accurate analysis of the $B$-$pB$ regression. 
The segments that contain $<$4 polarized brightness measurements were not considered in our analysis. 
By fitting the $B$-$pB$ regression with a linear function (Eq. (\ref{eq:linfun})), we obtained parameters $a_{\rm seg}$ and $b_{\rm seg}$ for each segment.

We excluded the segments with low correlation $\rho_{B,pB}<0.85$, since in these cases the linear model does not provide a satisfactory representation of the data. In addition, we excluded the segments with $a_{\rm seg} < 0.95$, which provide high values of $|C|$ (as explained in Sect. \ref{sec:3.2.1}) and, therefore, a biased estimate of $F$.

As shown in Sect. \ref{sec:3.2.1}, if the correlated variations of $B$ and $pB$ seen in a segment are due to the passage of the rotating streamers (as in simulations), the contribution of $C$ is not dominant, and, therefore, the best-fitting parameter $b_{\rm seg}$ can be considered as a reasonable estimate of the intensity of the F-corona. Moreover, if the F-corona is stable on the one-month timescale (as we assumed before), the distribution of $b_{\rm seg}$ values calculated for different segments over one month is expected to have a peak at the value of $F$. On the other hand, more complicated cases, such as passages of equatorial and/or broad streamers, provide the values of $C$ that deviate significantly from 0. This is expected to cause the broadening of $b_{\rm seg}$-distribution without shifting the peak.

\begin{figure*}
        \centering
        \includegraphics[width=0.9\textwidth]{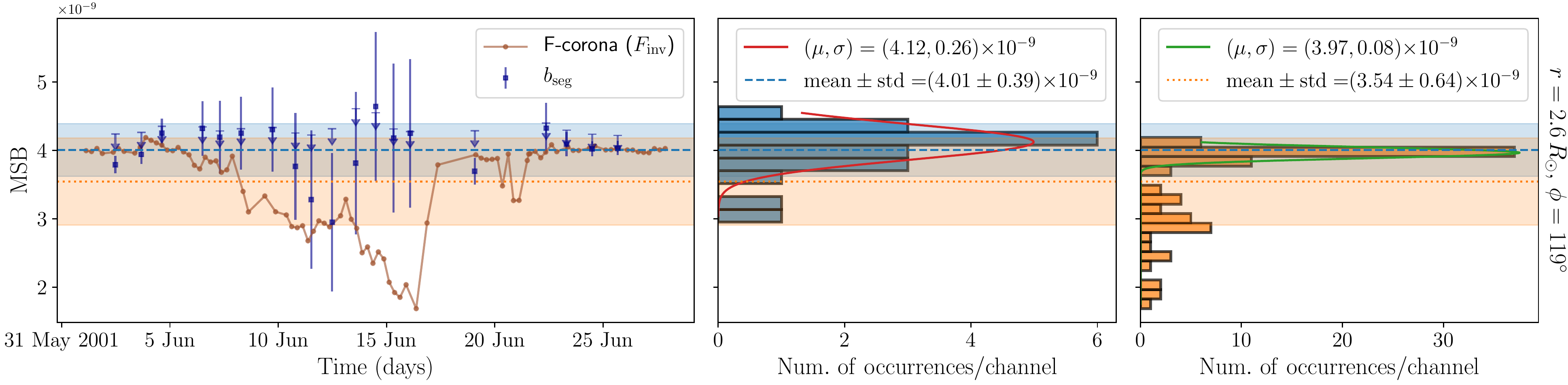} 
        \includegraphics[width=0.9\textwidth]{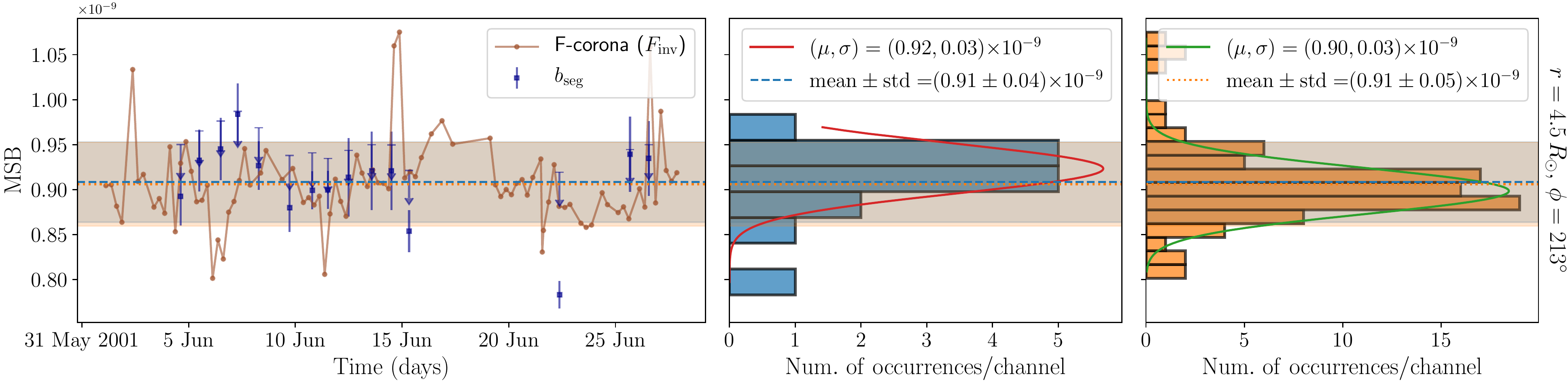}       
        \includegraphics[width=0.9\textwidth]{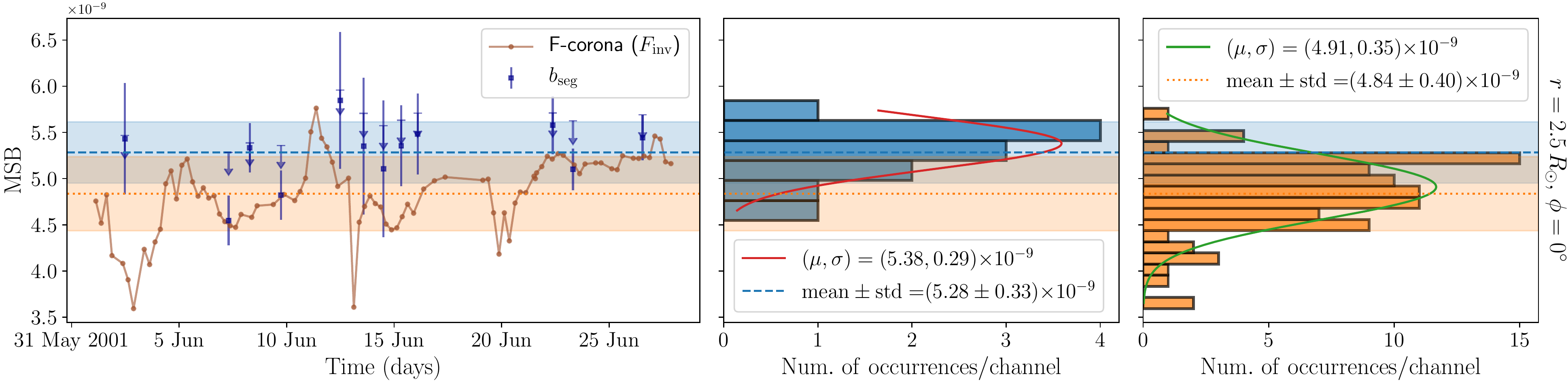} 
        \includegraphics[width=0.9\textwidth]{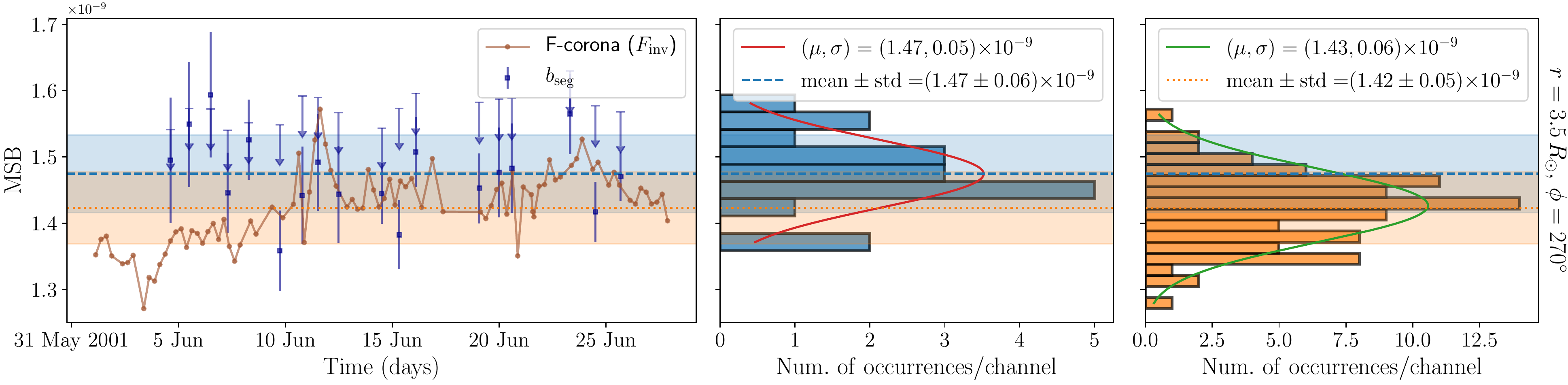} 
        \caption{Distribution of $b_{\rm seg}$ values (blue) and that of the F-corona intensity $F_{\rm inv}$ (orange) measured in different pixels in June 2001. The four rows correspond to pixels at $r=2.6$ $R_\sun$ and $\phi=119\degr$, $r=4.5$ $R_\sun$ and $\phi=213\degr$, $r=2.5$ $R_\sun$ and $\phi=0\degr$, and $r=3.5$ $R_\sun$ and $\phi=270\degr$ (from top to bottom). $F_{\rm inv}$ is calculated using the inversion method. The left column shows the evolution of both quantities. The mid and right columns show the histogram of the $b_{\rm seg}$ and $F_{\rm inv}$ distributions, respectively. In all panels, the $y$-axis is in the MSB units. 
        The red and green solid lines show the Gaussian fits of the $b_{\rm seg}$ and $F_{\rm inv}$ distributions, respectively. The mean and standard deviation of these distributions are shown by means of the blue dashed and orange dotted lines with the corresponding shaded area, respectively. 
        The upper limit of the F-corona intensity and the estimate of its systematic uncertainty inferred from the correlation method are shown for each segment as the blue arrow and the error bar, respectively (see text for details).}
         \label{fig:bseg_Finv_distr_all}
\end{figure*}

In our analysis, we considered overlapping segments: the beginning of each segment is shifted by one day with respect to the beginning of the previous segment (i.e., 2 days of overlap). This allowed us to increase the statistics of the $b_{\rm seg}$-distribution without significantly shifting its peak, since the segments with low correlation and $a_{\rm seg} < 0.95$ are excluded. In most cases, we obtained the $b_{\rm seg}$-distribution of a Gaussian shape (see some examples in Fig. \ref{fig:bseg_Finv_distr_all}).

The intensity of the F-corona ($F_{\rm corr}$) was calculated as the mean value of $b_{\rm seg}$-distribution. The corresponding statistical uncertainty ($\sigma_{\rm stat,~corr}$) was estimated as the spread of $b_{\rm seg}$ values (standard deviation). 
We also calculated the systematic uncertainty on $F_{\rm corr}$ caused by the lack of knowledge of the parameter $C$. We assumed that 20\% of $pB_{\rm max}$ in each segment can be considered as a reasonable estimate of the systematic uncertainty of the F-corona intensity. The final systematic uncertainty on $F_{\rm corr}$ ($\sigma_{\rm sys,~corr}$) is obtained by averaging the systematic uncertainties of all segments. Finally, we calculated an upper limit of the F-corona intensity equal to $(B-pB)$, and checked that the estimated value of $F_{\rm corr}$ satisfies this limit both during each single segment and the whole month.

\section{Results}\label{sec:4}
\subsection{F-corona images}\label{sec:4.1}
\begin{figure*}
        \centering
        \includegraphics[width=0.99\textwidth]{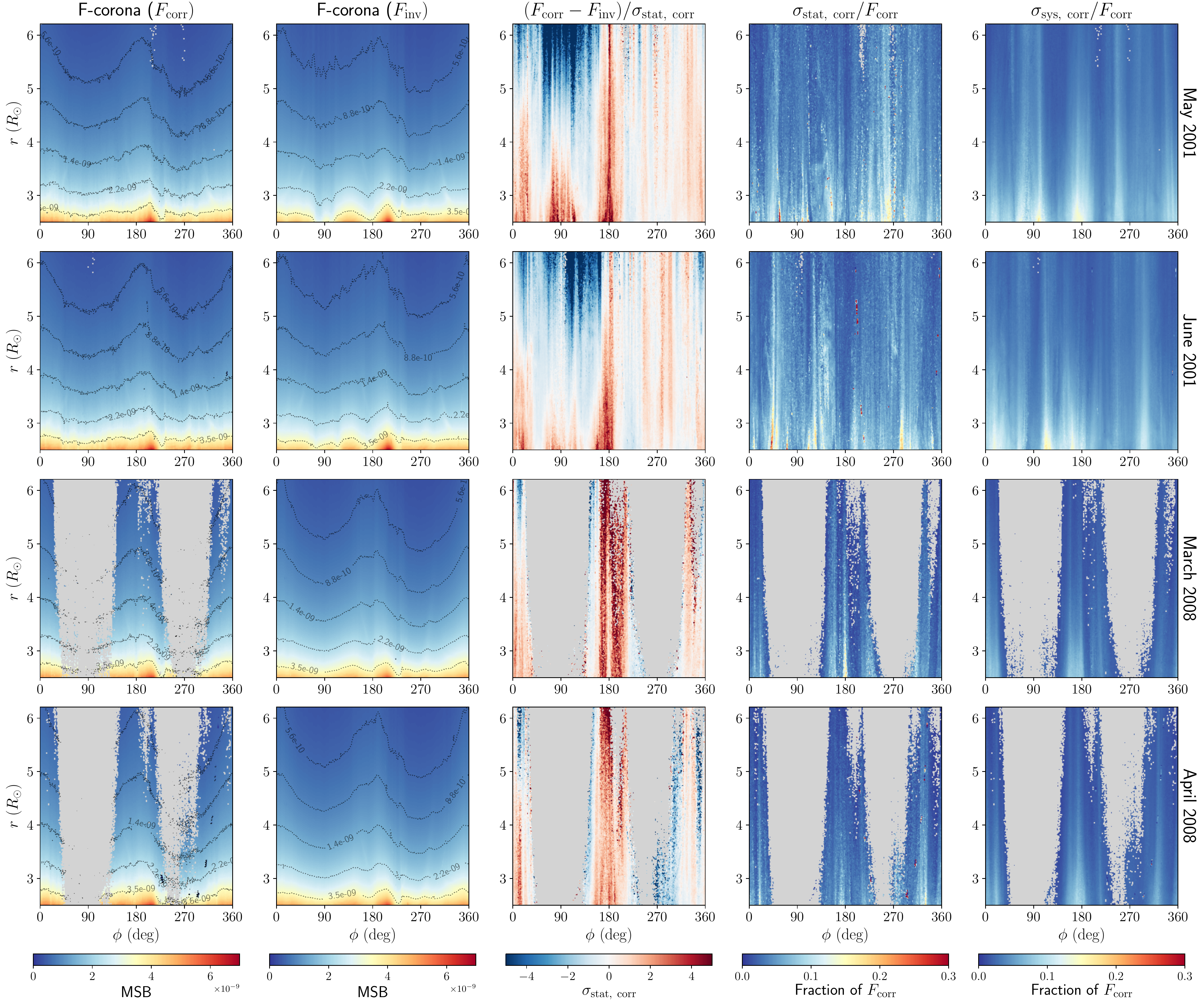} 
        \caption{F-corona maps calculated applying the correlation and inversion techniques for different months of data. The four rows correspond to May 2001, June 2001, March 2008, and April 2008 (from top to bottom). The first two columns show the intensity maps obtained with the correlation method ($F_{\rm corr}$) and the inversion method ($F_{\rm inv}$). The third column shows the difference maps $(F_{\rm corr} - F_{\rm inv})$ 
        normalized to the statistical uncertainties on $F_{\rm corr}$ ($\sigma_{\rm stat,~corr}$). The relative statistical and systematic uncertainties on $F_{\rm corr}$ are shown in the fourth and fifth columns, respectively. In all panels, the horizontal axis corresponds to the polar angle $\phi$ in degrees, and the vertical axis is the heliocentric distance $r$ in the $R_\sun$ units. The color bars represent the MSB in the first two columns, the $\sigma_{\rm stat,~corr}$ in the third column and the fraction of $F_{\rm corr}$ in the last two columns. The contour levels (dotted black lines) are reported in the MSB units. Blank pixels colored in gray appear in the regions where the correlation method is not applicable (see text for details).}
        \label{fig:Fmaps_no_Jan}
\end{figure*}

We repeated the procedure described in Sect. \ref{sec:3.2.2} for every pixel of the LASCO-C2 images and obtained 
the F-corona ($F_{\rm corr}$) intensity maps for two months during the solar maximum 
(May and June 2001) and two months during the solar minimum (March and April 2008). 
We calculated the maps that represent the relative statistical and systematic uncertainties on $F_{\rm corr}$. Both quantities typically do not exceed 10-15\%. Resulting maps are shown in Fig. \ref{fig:Fmaps_no_Jan}. 
The pixels with no measurement of $b_{\rm seg}$ were left blank. In addition, the pixels of the $\sigma_{\rm stat,~corr}$ maps for which the $b_{\rm seg}$-distribution contains only one measurement were left blank.

We also reconstructed the F-corona maps obtained through the inversion method ($F_{\rm inv}$) by averaging their evolution over a month and compared them with the $F_{\rm corr}$ maps. 
As shown in Fig. \ref{fig:Fmaps_no_Jan}, 
the brightness distribution obtained with the correlation method is smoother and does not contain the oversubtraction/undersubtraction features present in the $F_{\rm inv}$ maps (compare intensity contours in Fig. \ref{fig:Fmaps_no_Jan}, first and second columns). The most prominent features of $F_{\rm inv}$ maps can be seen in Fig. \ref{fig:Fmaps_no_Jan} (second column) at $\phi$ equal to 0\degr, 90\degr, and 180\degr. We calculated the difference between $F_{\rm corr}$ and $F_{\rm inv}$ normalized to the statistical uncertainties on $F_{\rm corr}$ ($\sigma_{\rm stat,~corr}$), and we found that in some regions it reaches significant values of up to 5-10 $\sigma_{\rm stat,~corr}$ (see Fig. \ref{fig:Fmaps_no_Jan}, third column). 
Such a difference is 
equivalent to $\sim$$10^{-9}$ MSB at low heliocentric distances. 

\begin{figure*}
        \centering
        \includegraphics[width=0.65\textwidth]{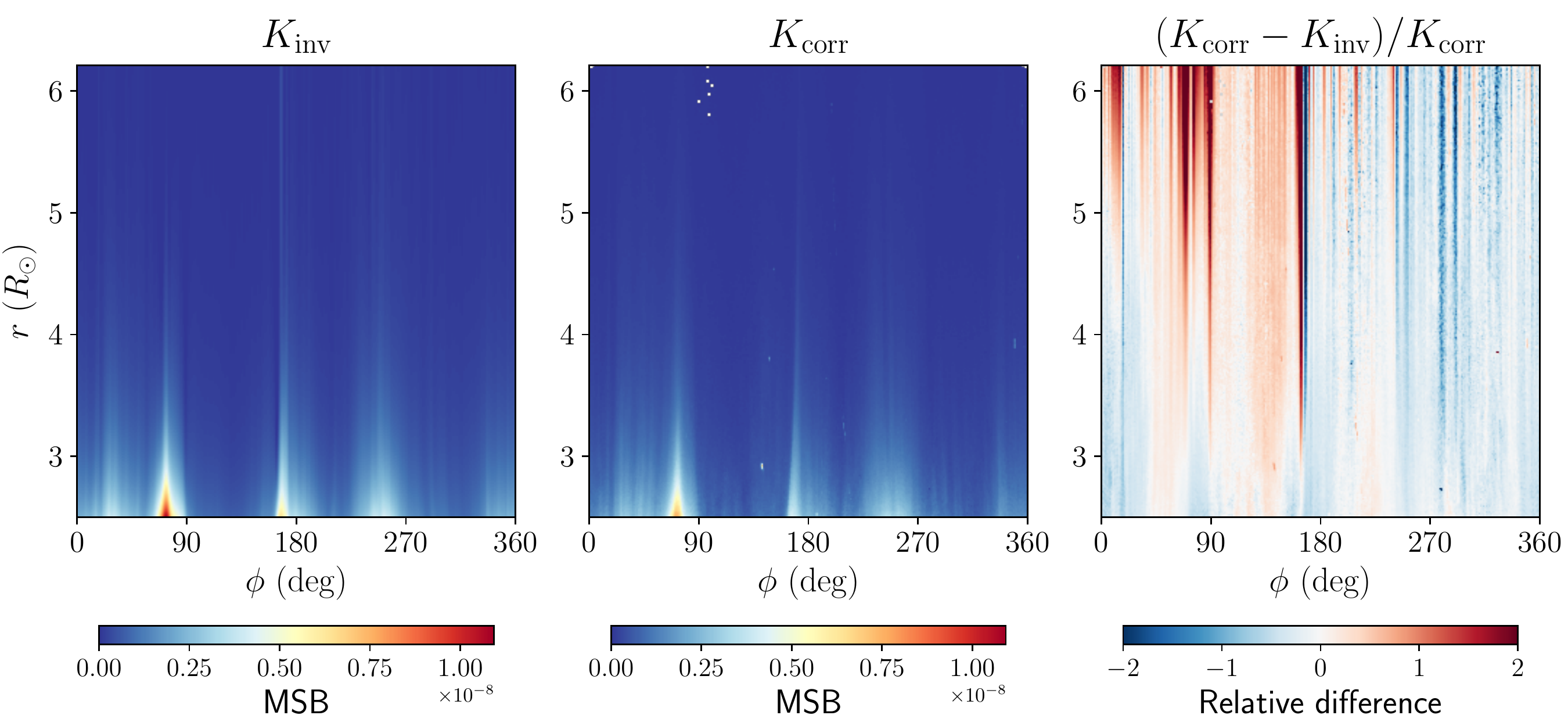} 
        \caption{K-corona maps calculated using the inversion ($K_{\rm inv}$, left panel) and correlation ($K_{\rm corr}$, mid panel) methods. $K_{\rm inv}$ is obtained using the $pB$ image acquired on June 1, 2001 at 9:00 and $B$ image on June 1, 2001 at 09:07 (as in Fig. \ref{fig:inv}). $K_{\rm corr}$ is calculated as the difference between the same total brightness image and the $F_{\rm corr}$ map obtained for June 2001. Right panel represents the relative difference between $K_{\rm corr}$ and $K_{\rm inv}$. The horizontal axis corresponds to the polar angle $\phi$ in degrees, and the vertical axis is the heliocentric distance $r$ in the $R_\sun$ units.}
        \label{fig:Kmaps_diffs}
\end{figure*}

Using the F-corona map obtained for June 2001, we reconstructed the image of the K-corona ($K_{\rm corr}$). For this, we subtracted $F_{\rm corr}$ from the total brightness image acquired on June 1, 2001 at 09:07. We compared $K_{\rm corr}$ with $K_{\rm inv}$ from Fig. \ref{fig:inv}, which was obtained using the same $B$ image and the preceding $pB$ image acquired on June 1, 2001 at 9:00. As shown in Fig. \ref{fig:Kmaps_diffs}, the relative difference between these two maps reaches significant values (0.5-5) in the regions where streamers appear.

\begin{figure}
        \centering
        \includegraphics[width=0.5\textwidth]{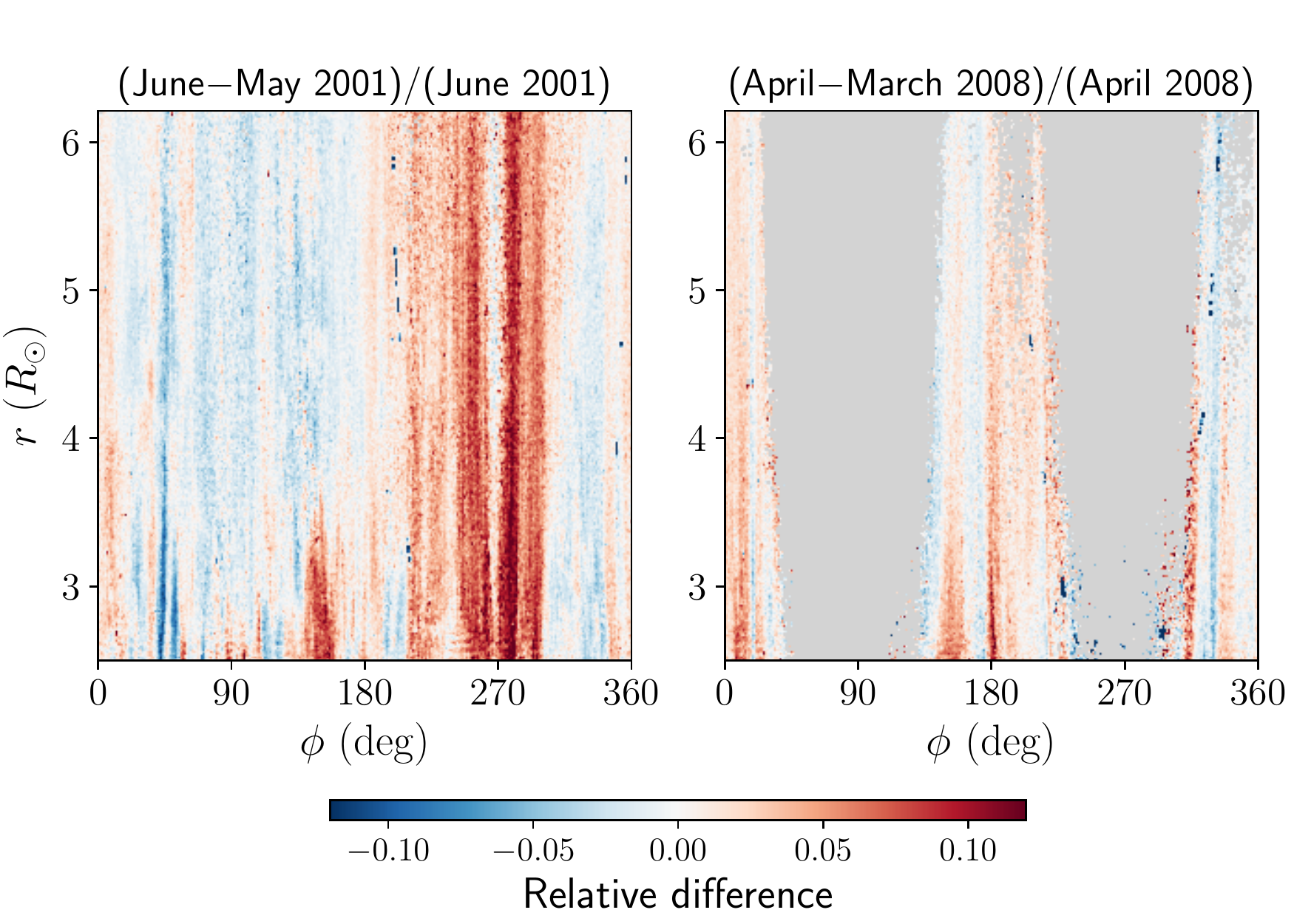} 
        \caption{Relative difference between $F_{\rm corr}$ maps calculated for two consecutive months. The horizontal axis corresponds to the polar angle $\phi$ in degrees, and the vertical axis is the heliocentric distance $r$ in the $R_\sun$ units. Blank pixels colored in gray appear in the regions where the correlation method is not applicable (see text for details).}
        \label{fig:Fmaps_diffs}
\end{figure}

We checked whether there is any significant variation of the F-corona intensity from one month to another. The difference 
maps calculated for two consecutive months (Fig. \ref{fig:Fmaps_diffs}) show that the relative variations are below $\sim$12\% ($\lesssim5\times10^{-10}$ MSB), which is within the statistical uncertainties of the $F_{\rm corr}$ measurement.

\subsection{Radial profiles of the F-corona}\label{sec:4.2}
\begin{figure}
        \centering
        \includegraphics[width=0.5\textwidth]{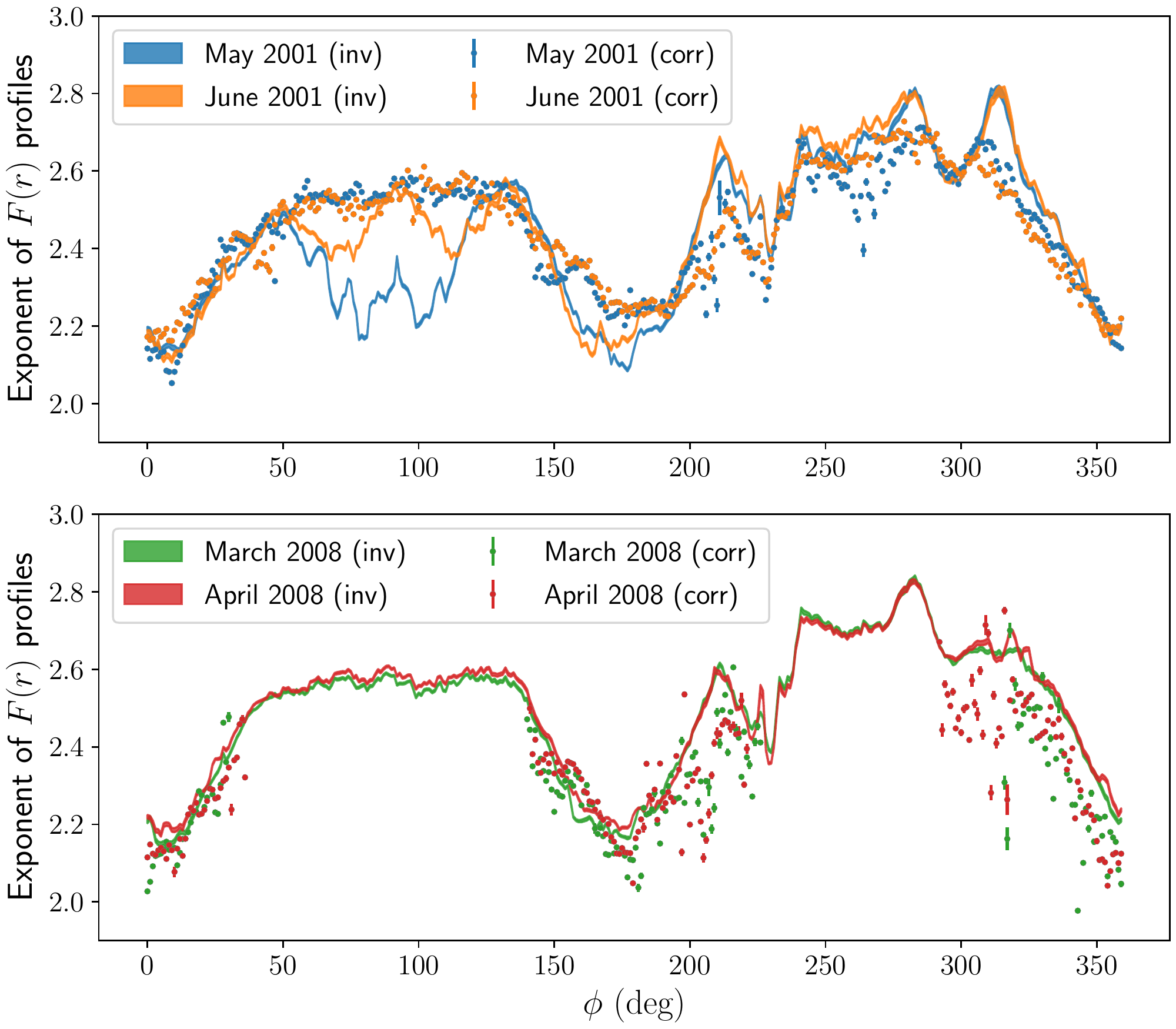} 
        \caption{Exponent of radial profiles of the F-corona calculated with the correlation (corr) and inversion (inv) techniques at $r\ge 3~R_\sun$. Top and bottom panels represent the values obtained for the solar maximum (May and June 2001) and solar minimum (March and April 2008) months, respectively. The best-fitting 1-$\sigma$ uncertainties are shown by means of the error bars for the correlation method and the width of shaded areas for the inversion method.}
        \label{fig:rad_grad}
\end{figure}

We compared the slope of radial profiles of the F-corona calculated with the correlation and inversion techniques at different polar angles. For this, we fit each profile $F(r)$ with a simple power-law function $f(r) \propto r^{-n}$ at $r\ge3~R_\sun$. Aiming to achieve a good quality of the fit while maintaining the simplicity of the fitting function, we did not consider the data between 2.5 $R_\sun$ and 3 $R_\sun$ where $F(r)$ changes its slope. The best-fitting exponent $n$ calculated for different months and polar angles is shown in Fig. \ref{fig:rad_grad}. We did not include the values of $n$ obtained for the polar regions (around 90$\degr$ and 270$\degr$) during the solar minimum. In these  quiet regions, less than a half of pixels is defined in the corresponding $F_{\rm corr}(r)$ profiles (for a detailed discussion, see Sect. \ref{sec:5.3}) and, as a consequence, the best fitting $n$ is not representative.

As can be seen from Fig. \ref{fig:rad_grad}, both the correlation and inversion methods provide a similar slope of $F(r)$ with $n\approx2.15$ measured along the equator and $n\approx2.6$ along the poles. These estimates are consistent with the values of $n$ derived using the intensity profiles from \citet{Saito1977} and close to those reported in \citet{KoutchmyLamy1985} for $r\ge4~R_\sun$. 

For the solar maximum months, the correlation method provides a smoother relation between $n$ and $\phi$ than the inversion method, especially at $\phi<200\degr$. The slopes of the $F_{\rm inv}(r)$ calculated around $90\degr$ and $180\degr$ for May and June 2001 deviate a lot from each other and from that of $F_{\rm corr}(r)$ due to the presence of prominent oversubtraction/undersubtraction features. The relation between $n$ and $\phi$ calculated for the solar minimum months is rather smooth for all cases, except for the March and April 2008 correlation measurements at $\phi$ around 200$\degr$ and 300$\degr$, where the F-corona image only partially defined. Finally, we point out that the F-corona profiles at $200\degr <\phi<320\degr$ are significantly affected by the contamination from the stray light (see Sect. \ref{sec:5.6} for discussion). The corresponding values of $n$ should be considered as indicative.

\section{Discussion}\label{sec:5} 
We presented a new correlation method for the direct measurement of the F-corona intensity from the VL data. We showed that the contribution from the F-corona can be reasonably well estimated in regions where the total and polarized brightness are highly correlated without any specific assumption on the geometry of the $n_{\rm e}$-distribution and/or the polarization of the K-corona.

\subsection{Simulation of streamer passages}\label{sec:5.1}
In order to estimate possible systematic uncertainties of the correlation method, we assumed that time variations of $B$ and $pB$ are mainly caused by passages of streamers which rotate solidly with the Sun. Following this assumption, we simulated how the quantities $K$ and $pB$ vary with time and obtained constraints on the parameter $C$. 

As mentioned in Sect. \ref{sec:3.2.1}, the streamers are simulated in the empty environment. 
However, the presence of electrons outside the streamers 
can introduce additional shift ($dC$) of the parameter $C$, which was not considered in our simulations. 
Although including this effect is beyond the presentation that we give in this paper, 
we nonetheless attempted to carry out a rough estimate of $dC$. For this, we assumed
that such an electron population has symmetric density distribution along the line LOS with respect to the POS. Applying the inversion method, we calculated the intensity of the K-corona ($K_{\rm inv}$) at the moment of time $t_{\rm min}$ when the minimum VL brightness is observed during a month. We also assumed that no streamer was passing in front of the observer at $t_{\rm min}$. We obtained that the normalized value of $|dC|$ calculated for each pixel of the image as in Eq. (\ref{eq:dC}) does not exceed $\sim$10\% of $pB_{\rm max}$: 
\begin{equation}
        dC = K_{\rm inv}(t_{\rm min}) - a\times pB(t_{\rm min}) \,,
        \label{eq:dC}
\end{equation}
where $a$ is the best fitting slope of the one-month $B$-$pB$ regression.

An accurate estimate of $dC$ and $C$ requires a more realistic modeling of the electron density distribution around the Sun (as, e.g., presented in \citealt{dePatoul2015}). 
However, our rough estimates suggest that the contribution from $dC$ should not shift the $F_{\rm corr}$ intensity calculated with the correlation method dramatically.

\subsection{Measuring the intensity of the F-corona}\label{sec:5.2}
For each pixel, we determined the intensity $F_{\rm corr}$ as the mean of $b_{\rm seg}$-distribution calculated for three-day segments of the VL observations. We compared the distribution of $b_{\rm seg}$ with that of the F-corona intensity ($F_{\rm inv}$) obtained by applying the inversion method to each pair of $pB$ and $B$ images.  
A number of the most representative cases are discussed below. 

In Fig. \ref{fig:bseg_Finv_distr_all} (first row), we show both distributions obtained at an intermediate polar angle ($\phi=119\degr$). 
The values of $b_{\rm seg}$ are rather stable, whereas the variation profile of $F_{\rm inv}$ is clearly anti-correlated with the $B$/$pB$ intensity (see also Fig. \ref{fig:pix_evo}). As mentioned in Sect. \ref{sec:3.1}, this happens due to the appearance of the asymmetric structures (such as streamers) in the solar corona during the period from June 7 through June 17, 2001, which leads to the oversubtraction of the K-corona. On the other hand, both methods provide similar estimates of the F-corona intensity outside this time interval, when there was no streamer passing through this region. 
Taking into account the spread of each distribution, the mean values $\langle b_{\rm seg} \rangle$ and $\langle F_{\rm inv} \rangle$ can be considered consistent with each other. We note, however, that $\langle F_{\rm inv} \rangle$ is lower than $\langle b_{\rm seg} \rangle$ due to the presence of the above-mentioned underestimated measurements. 
By removing such measurements from $F_{\rm inv}$ profiles, it is possible to obtain the results closer to those obtained with the correlation method. However, this would require pixel-by-pixel manipulation with the data.
An example for which both methods provide similar estimates of the F-corona intensity is shown in Fig. \ref{fig:bseg_Finv_distr_all} (second row, pixel at $\phi=213\degr$).

The equatorial and polar cases are shown in the last two rows of Fig. \ref{fig:bseg_Finv_distr_all}. In both examples, $F_{\rm corr}$ is higher than $\langle F_{\rm inv} \rangle$ due to presence of the oversubtraction features. The correlation method (if applicable) provides the most reliable results for the polar regions, where the value of $C$ is close to 0. However, the measurements of $F_{\rm corr}$ obtained for the equatorial regions can be affected by the underestimation of the parameter $|C|$.

The systematic uncertainties on $F_{\rm corr}$ defined as 20\% of $pB_{\rm max}$ are comparable with the standard deviation of $b_{\rm seg}$-distribution (comp. the blue error bars and the blue shaded area in Fig. \ref{fig:bseg_Finv_distr_all}).  
From this we conclude that the established values of the systematic uncertainties are not underestimated.

\subsection{Comparing the F-corona images}\label{sec:5.3}
\begin{figure}
        \centering
        \includegraphics[width=0.5\textwidth]{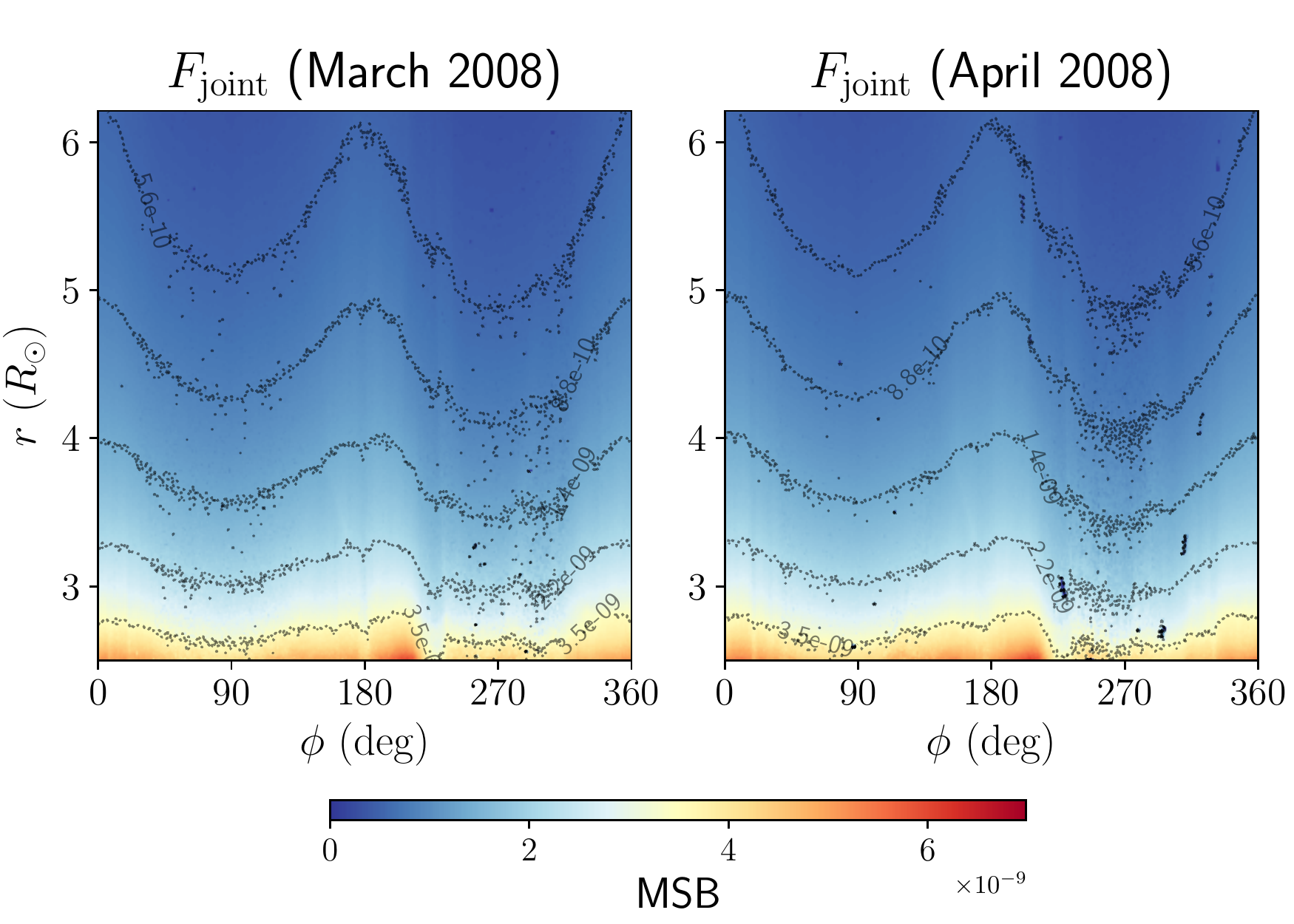} 
        \caption{Joint F-corona maps ($F_{\rm joint}$) calculated using the correlation method where applicable and the inversion method otherwise for March 2008 (left panel) and April 2008 (right panel).}
        \label{fig:Fmaps_2008_03_04_joint}
\end{figure}

\begin{figure}
        \centering
        \includegraphics[width=0.5\textwidth]{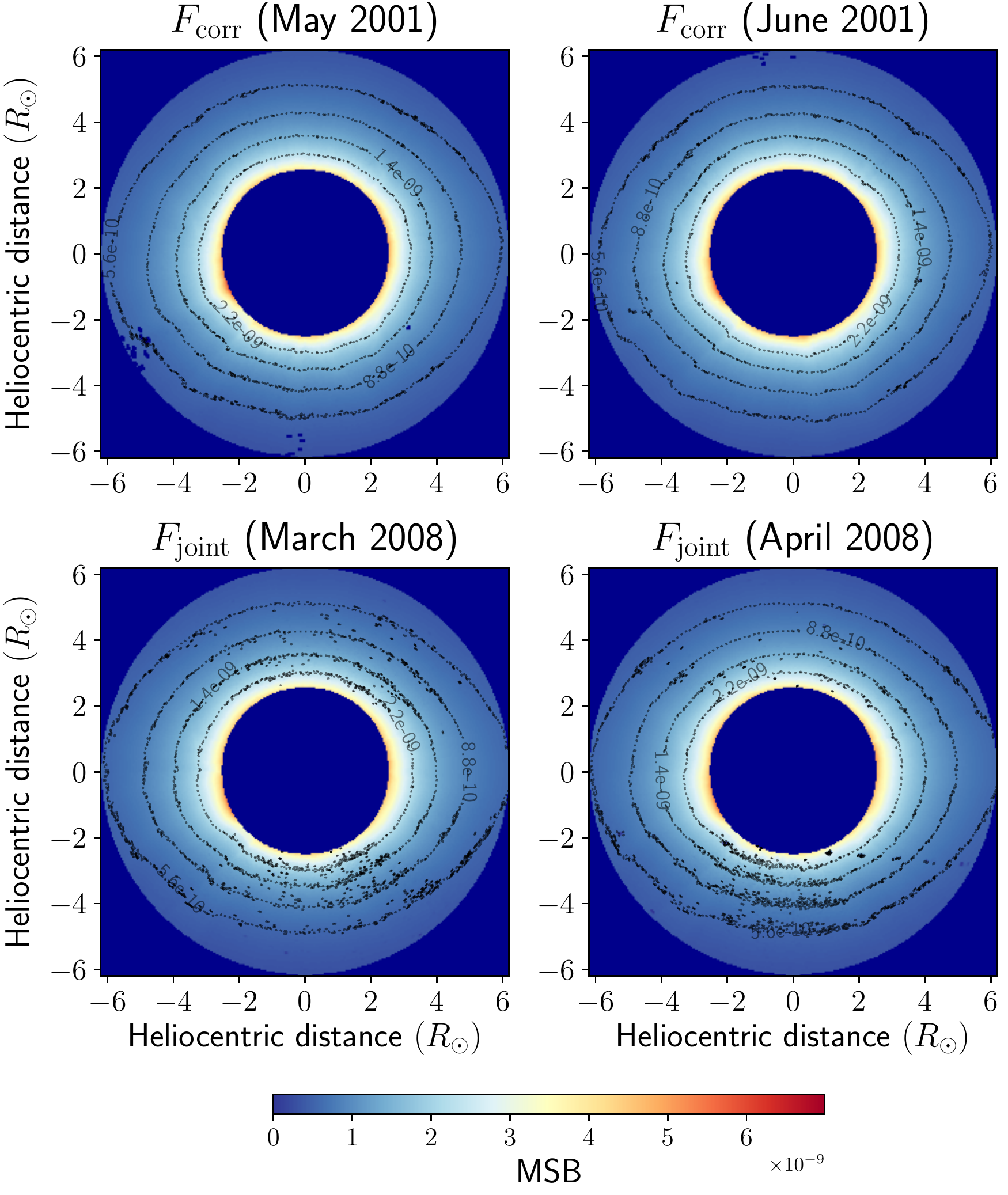} 
        \caption{$F_{\rm corr}$ and $F_{\rm joint}$ maps converted to Cartesian coordinates for May 2001 (top left panel), June 2001 (top right panel), March 2008 (bottom left panel), and April 2008 (bottom right panel). The color bar and contour levels represent the MSB units (as in Figs. \ref{fig:Fmaps_no_Jan} and \ref{fig:Fmaps_2008_03_04_joint}). In all panels, both axes correspond to the heliocentric distance in the $R_\sun$ units.}
        \label{fig:cart_Fmaps_no_Jan}
\end{figure}

Using the correlation method, we obtained the F-corona maps for different months during the solar maximum and solar minimum periods.
During solar maxima, the corona is active and it has a lot of streamers appearing at any polar angle. The total and polarized brightness are highly correlated in all regions, so the correlation method can be used for almost all pixels. During periods of minimum, the solar corona is rather quiet with relatively small number of streamers, which are located mainly in the equatorial region. The corresponding F-corona intensity maps have large areas of blank pixels at non-equatorial polar angles where the correlation $\rho_{B,pB}$ is low (see Fig. \ref{fig:corr_maps}). Although in these regions our method is not applicable, the missing values can be calculated by means of the inversion technique. In fact, since there is almost no streamer, the electron density can be considered symmetric, and $F_{\rm inv}$, therefore, can be considered as a reasonable estimate of $F$. We calculated the joint F-corona maps ($F_{\rm joint}$) using the correlation and inversion techniques in the high- and low- correlation regions, respectively (see Fig. \ref{fig:Fmaps_2008_03_04_joint}). The intensity contours of the resulting $F_{\rm joint}$ maps are rather smooth and do not show a significant step at the boundary of the two applied methods. This suggests that joining the correlation and inversion techniques it is possible to determine the complete F-corona intensity maps during the solar minimum. The final F-corona images converted back to Cartesian coordinates are shown in Fig. \ref{fig:cart_Fmaps_no_Jan}.

Compared to the inversion technique, the correlation method is particularly advantageous in the regions where bright streamers appear (especially for the months during the solar
maximum). In such regions, the difference between $F_{\rm corr}$ and $F_{\rm inv}$ can reach rather significant values up to $\sim$$10^{-9}$ MSB. 
As shown in the difference  
maps in Fig. \ref{fig:Fmaps_no_Jan}, for some polar angles (e.g., $\phi \sim 180\degr$) the electron density profiles $n_{\rm e}(r)$ calculated with the inversion technique provide the overestimated intensity (positive differences)  
of the K-corona $K_{\rm inv}$ at all heliocentric distances $r$ from 2.5 to 6.2 $R_\sun$. For other polar angles, for example around 0\degr and 90\degr, the $n_{\rm e}(r)$ profiles are such that the resulting intensity $K_{\rm inv}$ is overestimated at low ($r$$<$3--4 $R_\sun$, positive differences)  
and underestimated at high ($r$$>$4.5--5 $R_\sun$, negative differences)  
heliocentric distances.
Similar features are present in the difference map of the K-corona (Fig. \ref{fig:Kmaps_diffs}).

\subsection{Accounting for the presence of CMEs}\label{sec:5.4}
Apart from streamers, bright and dynamic K-corona events such as CMEs can cause the variation of the VL brightness. These events evolve much faster than streamers and produce sharp single-point peaks at the variation profiles of the total and polarized brightness (see e.g., a peak on June 20 in Fig. \ref{fig:pix_evo}). Given the cadence of the $pB$ acquisitions ($\sim$3 images per day) and different orientation of CMEs, 
each pixel of the $pB$ images converted to polar coordinates detects up to a few bright events (from 0 to $\sim$3) during a solar maximum month.

Similarly to narrow streamers, CMEs intersect the LOS in a very short time interval, so that in most cases the variation of $\alpha$ is expected to be small, $C$ close to 0, and the corresponding estimate of $F$ unbiased (see Sect. \ref{sec:3.2.1} and Appendix \ref{sec:app_alpha} for details). We found, however, that some CME measurements are outside an average trend of the $B$-$pB$ regression. This can be caused by the difference between the evolution of the geometry of the electron density distribution within a CME and simultaneously observed streamer(s). For example, if the ejected hot plasma is expanding along the LOS, the polarized fraction of the emitted radiation (and therefore the parameter $\alpha$) will constantly change and the parameter $C$ will reach significant values. In these cases, the corresponding $B$-$pB$ measurement can decrease the correlation coefficient of the segment of interest. 

The detailed characterization of the parameter $\alpha$ during a CME event is left for future studies. In this work, we estimated an error introduced by the presence of bright CMEs. Since it would be time consuming to manually remove each CME from the data, we implemented the following automatic procedure. Running our analysis, we examined the low correlation segments with $\rho_{B,pB}<0.85$ and calculated the correlation $\rho_i$ for each subsample of such a segment by omitting $i$-th observation. 
In case the correlation exceeded the threshold $\rho_i \ge 0.85$ for some iteration $i$, we stored the corresponding best fitting value of $b_{\rm seg}$ and calculated a new value of $F_{\rm corr}$.
The difference between new $F_{\rm corr}$ maps and those presented above is typically less than $\sim$0.46 $\sigma_{\rm stat,~corr}$ for solar maximum months. During the minimum phase of the solar cycle, CME events appear less frequently, and so the corresponding difference is expected to be even smaller.

\subsection{Accounting for the polarization of the F-corona}\label{sec:5.5}
We estimated the error introduced by neglecting the polarization $p_F(r)$ of the F-corona beyond 5 $R_\sun$. Following the model of \citet{Blackwell1966}, we assumed that $p_F$ is equal to 0 up to 5 $R_\sun$ and rises from 0.05\% at 5 $R_\sun$ to 0.16\% at 8 $R_\sun$ as a power-law function: $p_F \propto r^\gamma$. Taking into account that $pB$ is equal to $(pK + p_F F)$ and that the term $(p_F F)$ is constant in time, Eqs. (\ref{eq:linfun}) and (\ref{eq:Beq4}) can be rewritten as
\begin{equation}
        B = a \times pB + b \equiv a \times (pK + p_F F) + b \,,
        \label{eq:linfun_pF}
\end{equation}
\begin{equation}
        B(pK) = a \times pK + C +  F\,.
        \label{eq:Beq_pF}
\end{equation}
Assuming again that the contribution from $C$ is not significant, one can derive that
\begin{equation}
        F=b/(1-a p_F) \, .
        \label{eq:F_pF}
\end{equation}

By adjusting an estimate of $F$ in each three-day segment as shown in Eq. (\ref{eq:F_pF}), we calculated new values of $F_{\rm corr}$ for each pixel at heliocentric distances $>$5 $R_\sun$. They differ from $F_{\rm corr}$ values presented above 
by $\lesssim$0.13 $\sigma_{\rm stat,~corr}$ and $\lesssim$0.26 $\sigma_{\rm stat,~corr}$ for solar maximum and minimum months, respectively.

\subsection{Contamination from the stray light}\label{sec:5.6}
The F-corona maps presented in this work contain non-negligible contribution from the instrumental stray light. For example, one of its most prominent features arises from the occulter pylon and can be seen as a relatively faint sector at $\phi\approx230\degr$ (see Fig. \ref{fig:Fmaps_no_Jan}). 

\begin{figure}
        \centering
        \includegraphics[width=0.27\textwidth]{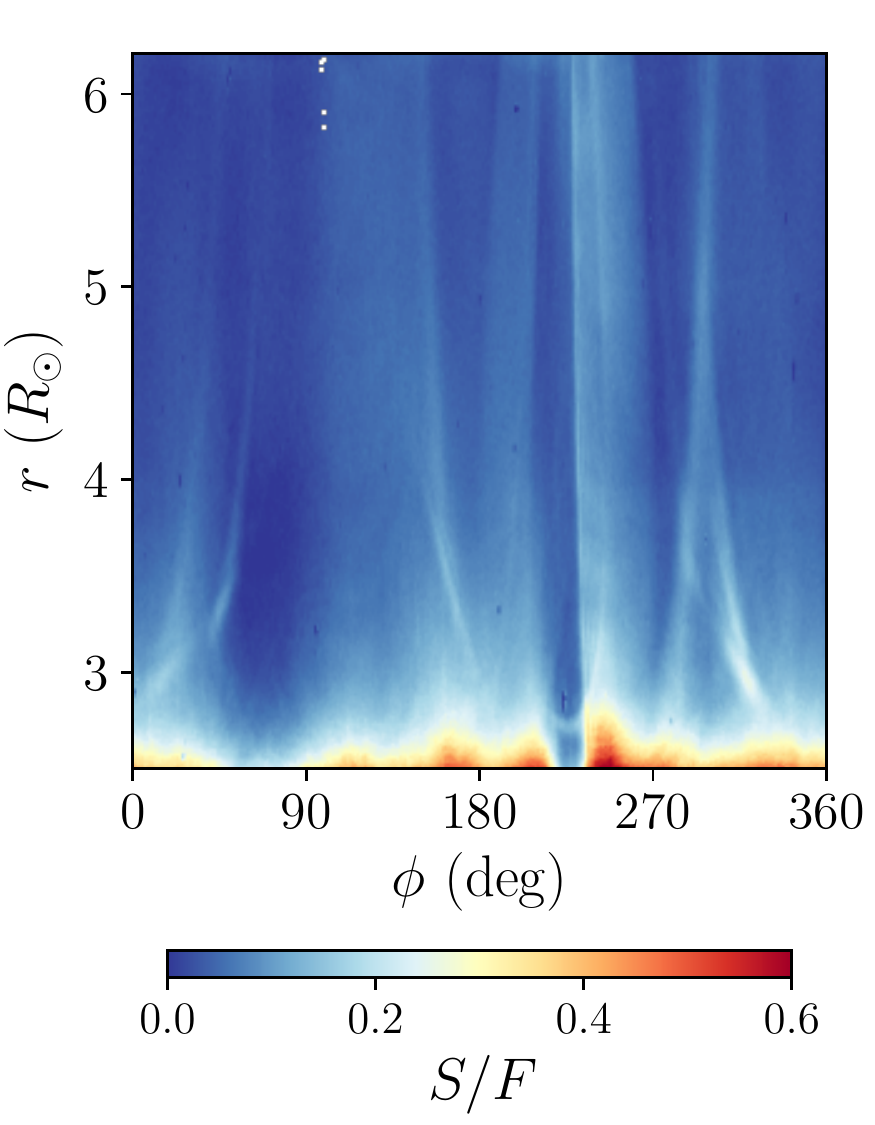}  
        \caption{Ratio map $(S/F)$ calculated for June 2001. The stray light pattern $S$ is taken from \citet{Llebaria2021_F_SL}. The F-corona map $F$ is calculated by applying the correlation method to the newly calibrated data presented by \citet{Lamy2020_C2, Llebaria2021_F_SL}.}
        \label{fig:S_F_ratio} 
\end{figure}

Disentangling the F and stray light components is a nontrivial task, which was beyond the original scope of this paper. Recently, \citet{Llebaria2021_F_SL} presented the sophisticated procedure for separating these two components from the LASCO-C2 data obtained over 24 years. They used recalibrated total and polarized brightness images (for details, see \citealt{Lamy2020_C2}) and restored 36 maps that account for the relatively slow temporal variation of the stray light pattern of the LASCO-C2 instrument. We could not use these patterns to remove the stray light from our $F_{\rm corr}$ and $F_{\rm inv}$ images, since they were obtained using the data set calibrated in a different way. In order to evaluate the contamination $S$ from the stray light, we repeated our analysis using the newly calibrated total and polarized brightness images stored in the corresponding data archive\footnote{\url{idoc-lasco-c2-archive.ias.u-psud.fr}}, as specified in \citet{Lamy2020_C2, Llebaria2021_F_SL}. As shown in Fig. \ref{fig:S_F_ratio}, the ratio $S/F$ ranges from $\sim$0 up to 0.5$-$0.6.

We also found that the recalculated $F_{\rm corr}$ maps are consistent with those restored by \citet{Llebaria2021_F_SL}. The relative differences typically do not exceed $\sim$10-15\%, which is comparable with the uncertainties of the correlation method.

\section{Conclusions}\label{sec:6}
The correlation method is a new model-independent and data-driven approach for calculating the intensity maps of the F-corona. It provides a more accurate estimate of $F$ than the inversion method in the active regions with streamers and/or CMEs: that is almost the full LASCO-C2 FOV during the solar maximum and about a half of it during the solar minimum. We point out that the correlation method does not require any specific assumption on the geometry of the $n_{\rm e}$-distribution (as required, e.g., for the inversion technique) and/or a specific model of the K-corona polarization (as required, e.g., for the analysis of \citealt{Lamy2020_C2, Llebaria2021_F_SL}).

We evaluated the uncertainties introduced by not considering the polarization of the F-corona beyond 5 $R_\sun$ and the presence of bright CMEs, and found that they are significantly lower than the statistical errors. As another confirmation of the reliability of the correlation method, we found that it provides results consistent with those of \citet{Llebaria2021_F_SL}, who presented one of the most accurate procedure for the restoration of the F-corona from the LASCO-C2 data. We also obtained that the slope of the radial profiles $F(r)$ calculated at different polar angles is in agreement with the values reported in the literature.

During the periods of solar minimum, our method is not applicable to the quiet polar regions with no streamers. In these parts of the LASCO-C2 FOV the inversion method is expected to provide a reliable estimate of $F$. Combining the correlation and inversion methods, it is possible to obtain the complete intensity maps of the F-corona during the solar minimum. 

Another limitation of the correlation method is that the analyzed data set should cover a certain period of time with a reasonable cadence for measuring the correlation of the total and polarized brightness and determining an accurate estimate of $F$. As we showed above (e.g., in Fig. \ref{fig:corr_maps}), one month of data is definitely sufficient to detect the activity in the majority of pixels of the LASCO-C2 FOV during the solar maximum period and in the equatorial pixels during the solar minimum periods. The cadence of $\sim$3 measurements per day allowed us to achieve an accuracy of about 10-15\%. In this respect, the inversion technique is simpler to use, since the $F_{\rm inv}$ map can be derived from a pair of $B$ and $pB$ images. On the other hand, such F-corona images will be distorted by the oversubtraction/undersubtraction features even more than the monthly averaged $F_{\rm inv}$ maps shown in Fig. \ref{fig:Fmaps_no_Jan}. 

Resulting $F_{\rm corr}$ maps can be used for reconstructing accurate images of the K-corona, which is necessary to characterize the electron population within the solar corona. By applying the correlation method to larger data sets, it will be possible to analyze the time evolution of the F-corona brightness distribution observed with LASCO. Our method does not allow us to separate the contribution of the stray light from the F-corona. For this purpose, one can use the stray light patterns obtained by means of other techniques. Once corrected for the presence of the stray light, accurate $F_{\rm corr}$ images can be used for studying the distribution of the interplanetary dust.

Although the correlation method already provides reliable results, a number of further steps can be taken to improve it: 
\begin{enumerate}
        \item By developing more realistic simulations, it will be possible to establish more accurate constraints on $C$ and to decrease the corresponding systematic uncertainties on $F$. For example, one can consider more complex shapes of a streamer with the orientation evolving with time. Including a population of electrons outside the streamers, we will be able to evaluate $dC$ directly from the simulations. 
        \item The detailed investigation of the passages of equatorial and/or broad streamers is necessary for providing a more reliable estimate of the F-corona intensity for these cases.
        \item By investigating the variation of the polarized fraction of the K-corona (equiv. to $1/\alpha$) during the CME events, it will be possible to explain the presence of the CME measurements that are outside an average trend of the $B$-$pB$ regression. Carefully removing such measurements from the data will allow us to improve an accuracy of the $F_{\rm corr}$ maps.
        \item Additional improvements can be achieved accounting for the polarization of the F-corona above several $R_\sun$.
\end{enumerate}

With the high-cadence $pB$ observations, it will be possible to obtain a more accurate measurement of $F$, decreasing both the statistical and the systematic uncertainties. Such observations will be carried out in the near future with the Metis instrument \citep{Antonucci2020_metis} -- a coronagraph onboard the Solar Orbiter mission \citep{GarciaMarirrodriga2021_SolarOrbiter} that is currently in the cruise phase. In contrast to LASCO, Metis is going to acquire both the total and polarized brightness images during each VL observation, providing more frequent series of images for the correlation analysis. In addition, the out-of-ecliptic observations of Metis will allow us to obtain the longitudinal brightness distribution of the F-corona for the first time.

\begin{acknowledgements}
We thank the referee for his/her in-depth comments. A.B. would like to thank Dr. Polina Zemko for very useful discussions and suggestions. This activity has been supported by ASI and INAF under the contract: Accordo ASI-INAF \& Addendum N. I-013-12-0/1. In this work we made use of the following \texttt{Python} packages: \texttt{Matplotlib} \citep{Matplotlib2007}, \texttt{NumPy} \citep{2020NumPy-Array}, \texttt{SciPy} \citep{2020SciPy-NMeth}, \texttt{Astropy} \citep{Astropy2013, Astropy2018}. This work makes use of the LASCO-C2 legacy archive data produced by the LASCO-C2 team at the Laboratoire d'Astrophysique de Marseille and the Laboratoire Atmosph\`eres, Milieux, Observations Spatiales, both funded by the Centre National d'Etudes Spatiales (CNES). LASCO was built by a consortium of the Naval Research Laboratory, USA, the Laboratoire d'Astrophysique de Marseille (formerly Laboratoire d'Astronomie Spatiale), France, the Max-Planck-Institut f\"ur Sonnensystemforschung (formerly Max Planck Institute f\"ur Aeronomie), Germany, and the School of Physics and Astronomy, University of Birmingham, UK. SOHO is a project of international cooperation between ESA and NASA.
\end{acknowledgements}

\begin{appendix}

\section{Deriving the relation between the parameters $\alpha$, $a,$ and $C$}\label{sec:app_alpha}
Assuming that $pB_0$ is the polarized brightness at time $t_0$ and that $\alpha_0=\alpha(pB_0)$ as boundary conditions, we integrated the differential Eq. (\ref{eq:dalpha}) and obtained the following:
\begin{equation}
        \ln\frac{pB}{pB_0} = - \ln\frac{a-\alpha(pB)}{a-\alpha_0} \, ,
        \label{eq:app_alpha1}
\end{equation}
or, alternatively,
\begin{equation}
        \frac{pB_0}{pB} = \frac{a-\alpha(pB)}{a-\alpha_0} \,.
        \label{eq:app_alpha2}
\end{equation}

From Eq. (\ref{eq:app_alpha2}), we derived the solution of the differential equation in a general form:
\begin{equation}
        \alpha(pB) = a - \frac{pB_0\,(a-\alpha_0)}{pB}.
        \label{eq:app_alpha_sol}
\end{equation} 
The simplified form of this equation was obtained 
assigning the expression $pB_0\,(\alpha_0-a)$ to the parameter $C$ (see Eq. (\ref{eq:dalpha_sol})).
In addition, from Eq. (\ref{eq:app_alpha_sol}) we can find that 
\begin{equation}
        \alpha(pB)\,pB = K(pB) = a\,pB + C
        \label{eq:lin_fun_2_ponts}
\end{equation}
and
\begin{equation}
        a = \frac{\alpha\,pB - \alpha_0\,pB_0}{pB - pB_0} = \frac{K - K_0}{pB - pB_0}
        \label{eq:a_slope_2_points}
.\end{equation}

\begin{figure}
        \centering
        \includegraphics[width=0.5\textwidth]{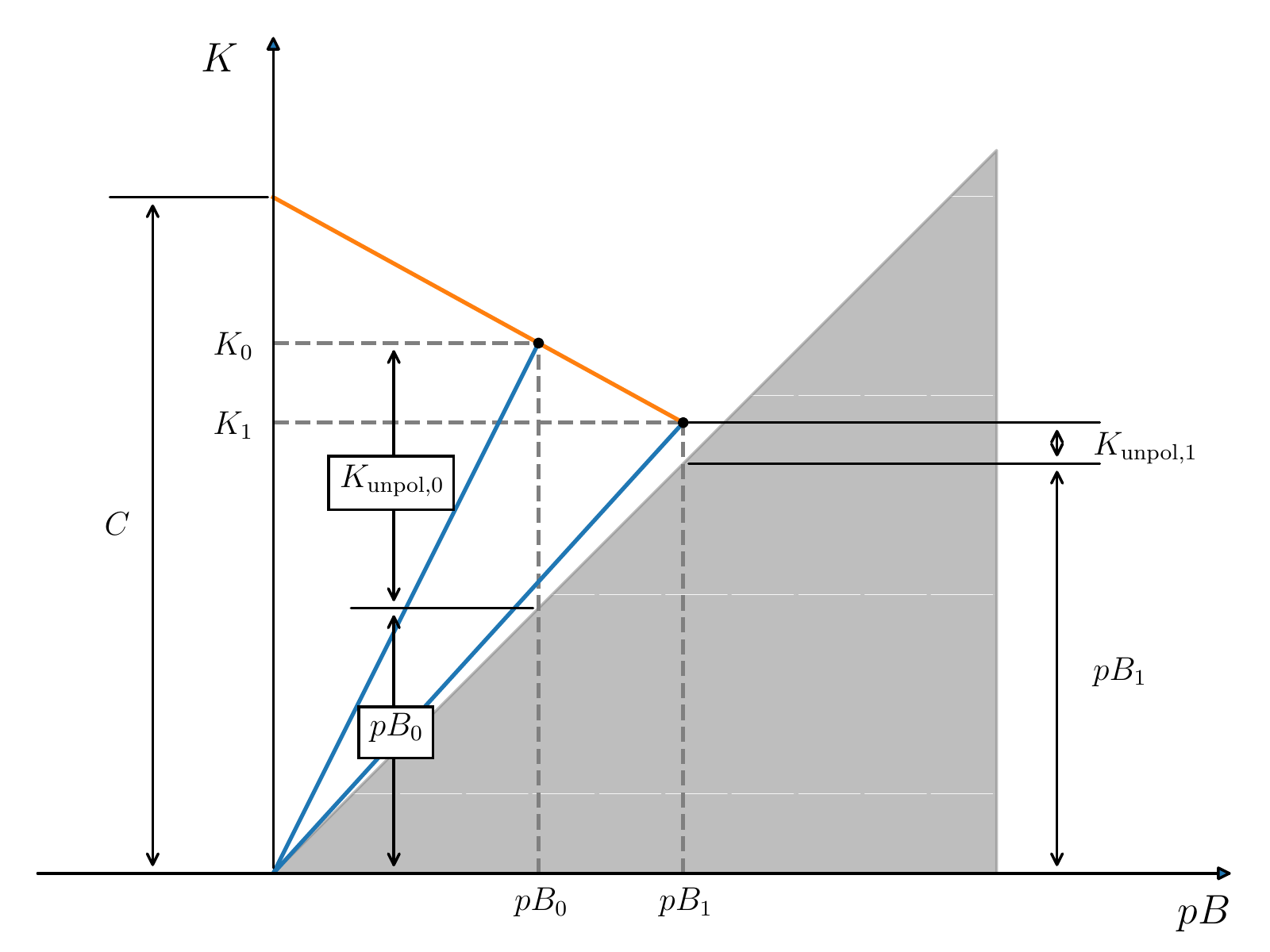} 
        \caption{Two consecutive measurements of $K$ and $pB$ at time $t_0$ and $t_1$ (black dots). The orange solid line with a slope equal to $a$ passes through both points and intersects the $y$-axis at $K=C$. Since $K \ge pB$, no ($pB$, $K$)-point can appear within the gray shaded area below the line $K=pB$ with a slope equal to 1. $K_{\rm unpol}$ corresponds to the unpolarized fraction of the intensity $K$. }
        \label{fig:K_pB_sketch}
\end{figure}

In Fig. \ref{fig:K_pB_sketch}, we show a schematic plot with two measurements of $K$ and $pB$ performed in close moments of time ($t_0$ and $t_1$). The linear function which passes thorough both points is given by Eq. (\ref{eq:lin_fun_2_ponts}), and has a slope equal to $a$, as can easily be derived from Eq. (\ref{eq:a_slope_2_points}). It intersects the $y$-axis ($pB=0$) at the value equal to $C$ (see Eq. (\ref{eq:lin_fun_2_ponts})). 
The parameter $C$ can thus be interpreted as the value reached by intensity $K$ when its polarized fraction is equal to 0 and the slope of the $K$-$pB$ regression is equal to $a$. 

\begin{figure}
        \centering
        \includegraphics[width=0.24\textwidth]{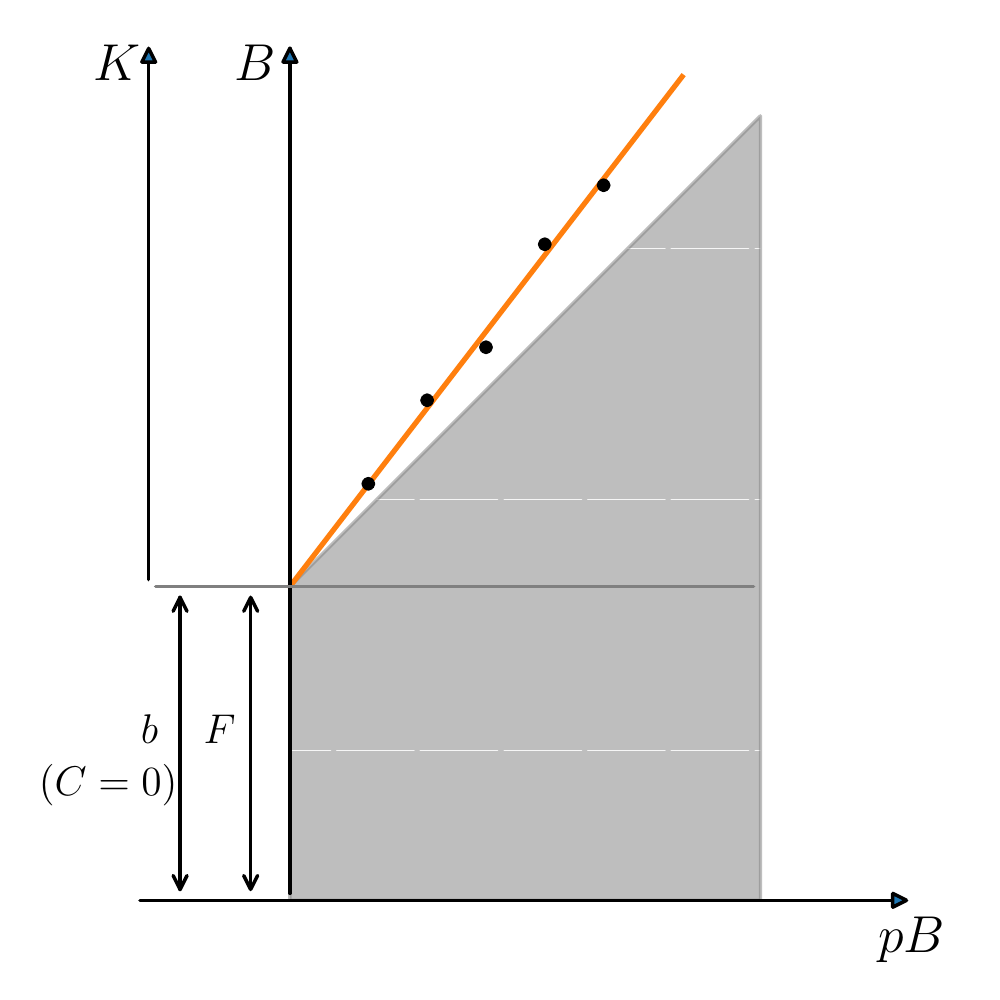} 
	
        \includegraphics[width=0.24\textwidth]{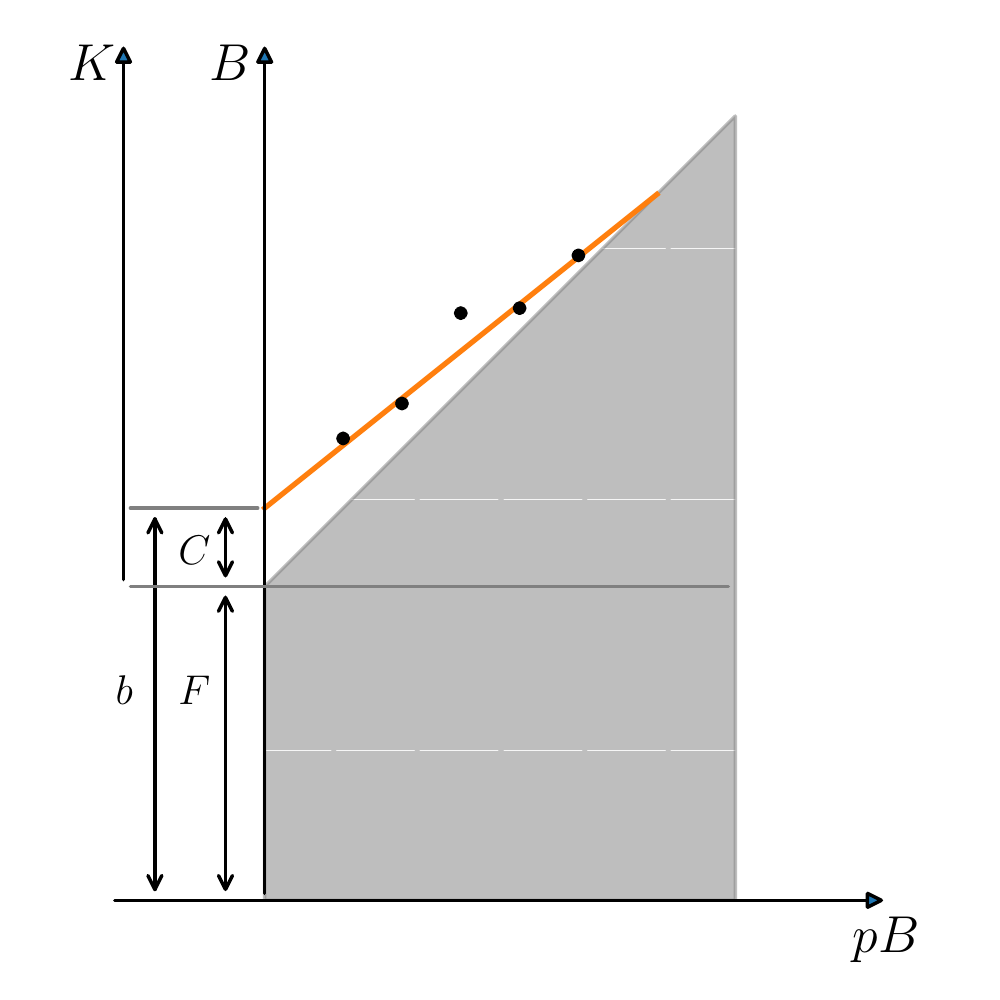}
        \includegraphics[width=0.24\textwidth]{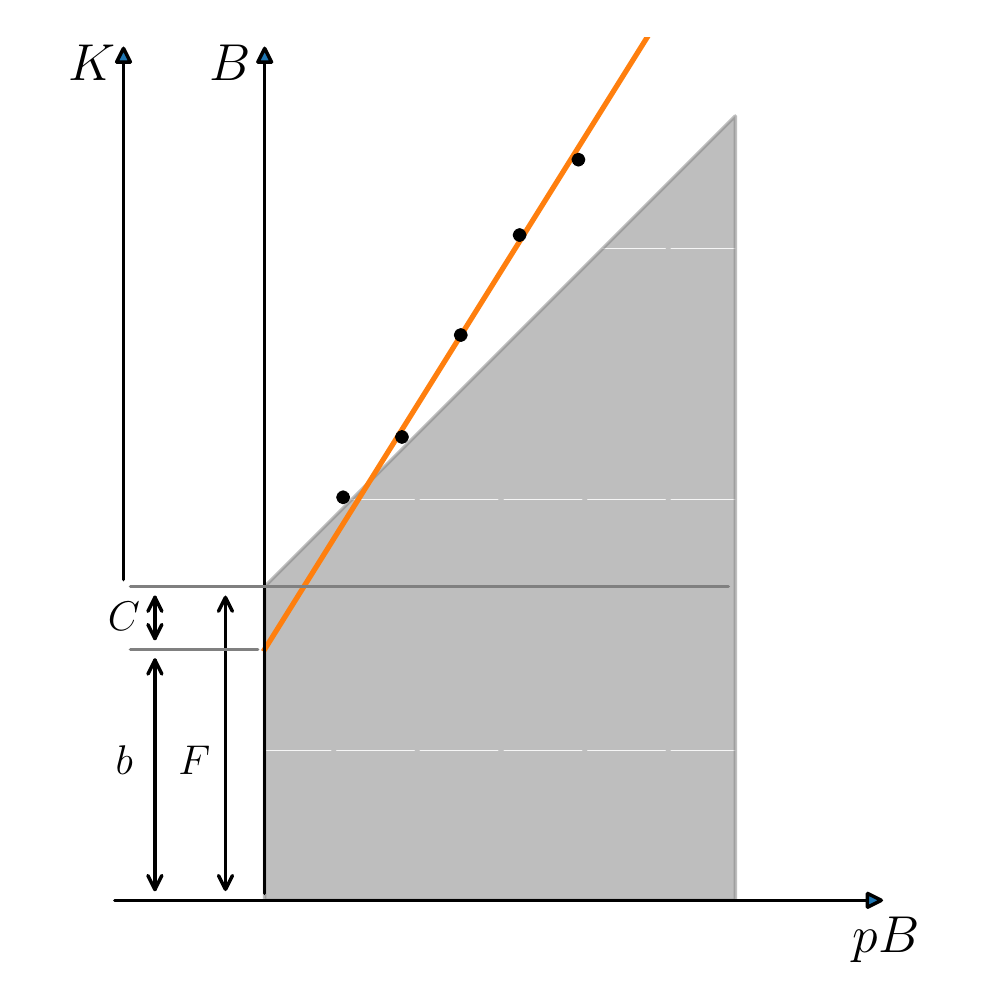}
        \caption{Linear function (orange line) fit to the $pB$, $B$-measurements (black dots). Three panels represent the cases when the best-fitting parameter $b$ provides an unbiased estimate of the F-corona intensity $F$ (top left panel, $C=0$), an overestimated value of $F$ (top right panel, $C>0$), and an underestimated value of $F$ (bottom panel, $C<0$). Since $K \ge pB$, no ($pB$, $B$)-point can appear within the gray shaded area below the line $B=pB+F$ with a slope equal to 1.}
        \label{fig:B_pB_3sketches}
\end{figure}

The total brightness images contain the contribution from both the K- and F- coronae, and, as mentioned in Sect. \ref{sec:3.2}, the parameters $C$ and $F$ cannot be derived from the linear fit of the $B$-$pB$ regression. In fact, the constant term $b$ of the best-fitting function can, in principle, provide either an unbiased estimate of the F-corona intensity (if $C=0$), an overestimated value of $F$ (if $C>0$), or, alternatively, an underestimated value of $F$ (if $C<0$). The schematic representation of these three cases is shown in Fig. \ref{fig:B_pB_3sketches}.

\section{Details on the streamer simulations}\label{sec:app_sim_tab_fig}
\begin{figure}
        \centering
        \includegraphics[width=0.5\textwidth]{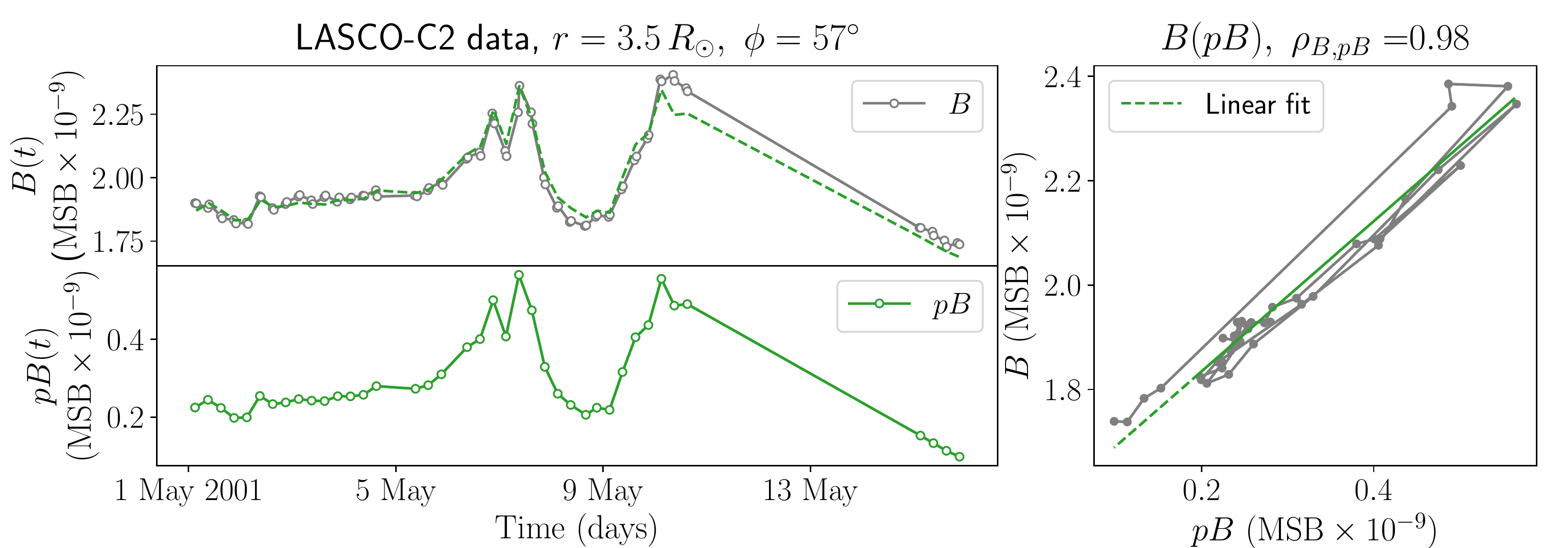} 
        \includegraphics[width=0.5\textwidth]{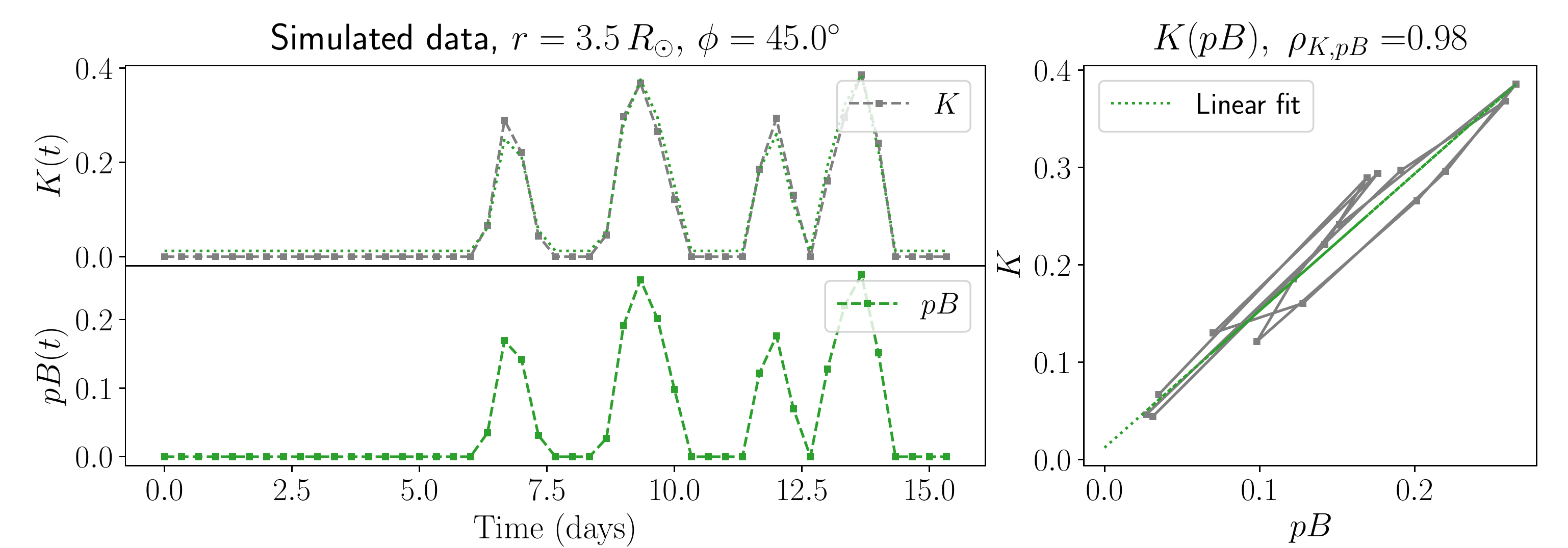} 
        \includegraphics[width=0.43\textwidth]{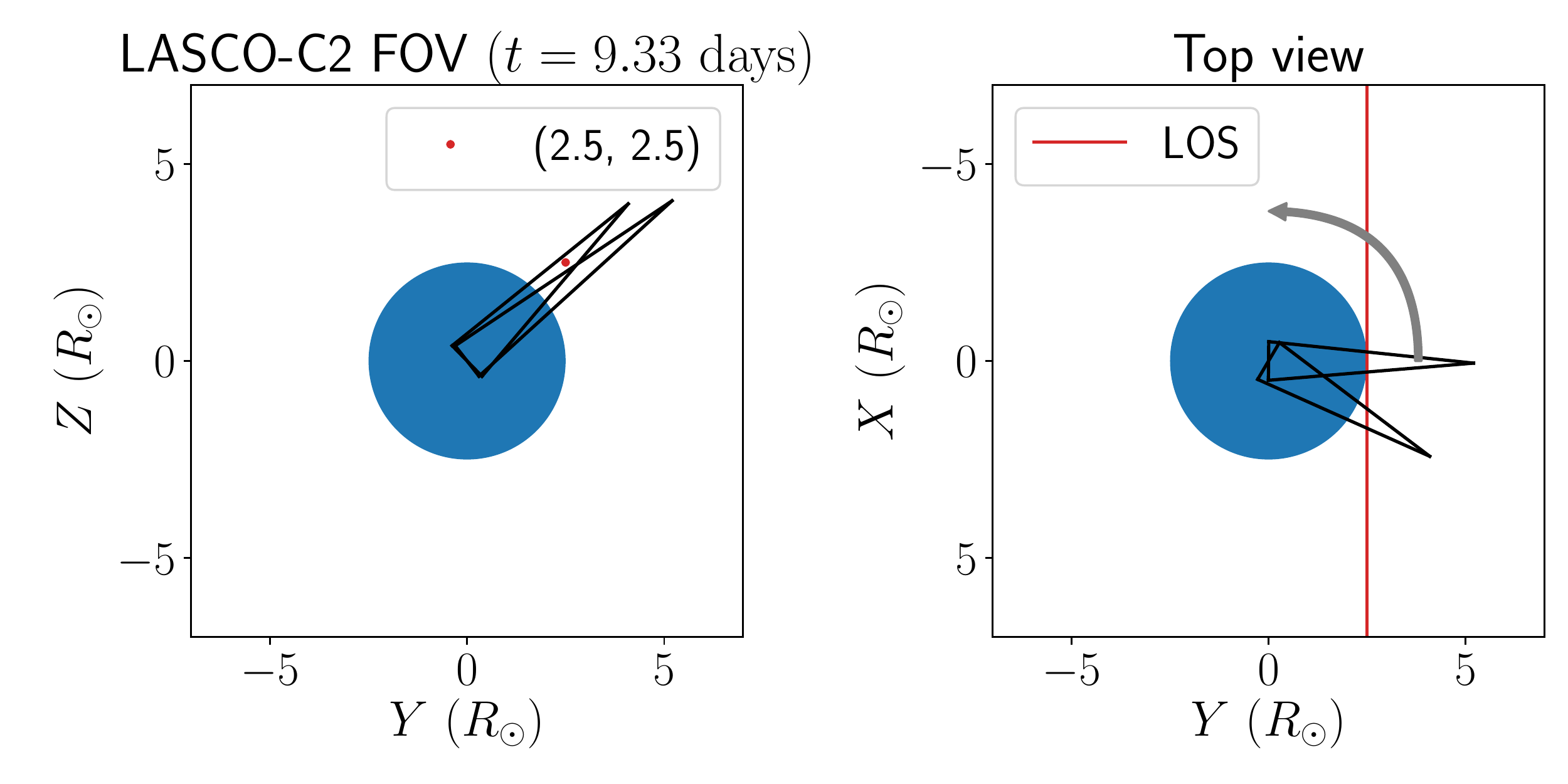} 
        \caption{Same as Fig. \ref{fig:sim_val_int1_broad}, but for the pixel at $r=3.5$ $R_\sun$ and $\phi\approx50\degr$ (May 2001). Bottom panels represent the orientation of two narrow simulated streamers at $t=9.33$ days, as defined in Table \ref{tab:all_sims} (``Sim2'').}
        \label{fig:sim_val_int2_narrow}
\end{figure}

\begin{figure}
        \centering
        \includegraphics[width=0.5\textwidth]{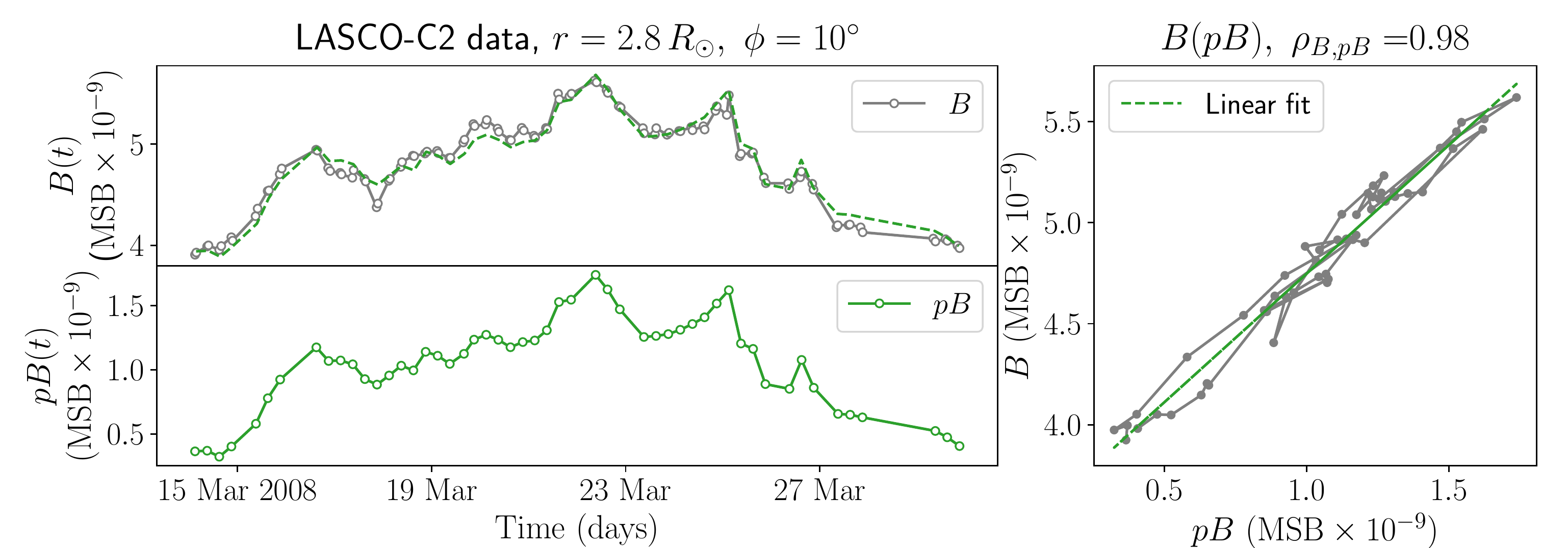} 
        \includegraphics[width=0.5\textwidth]{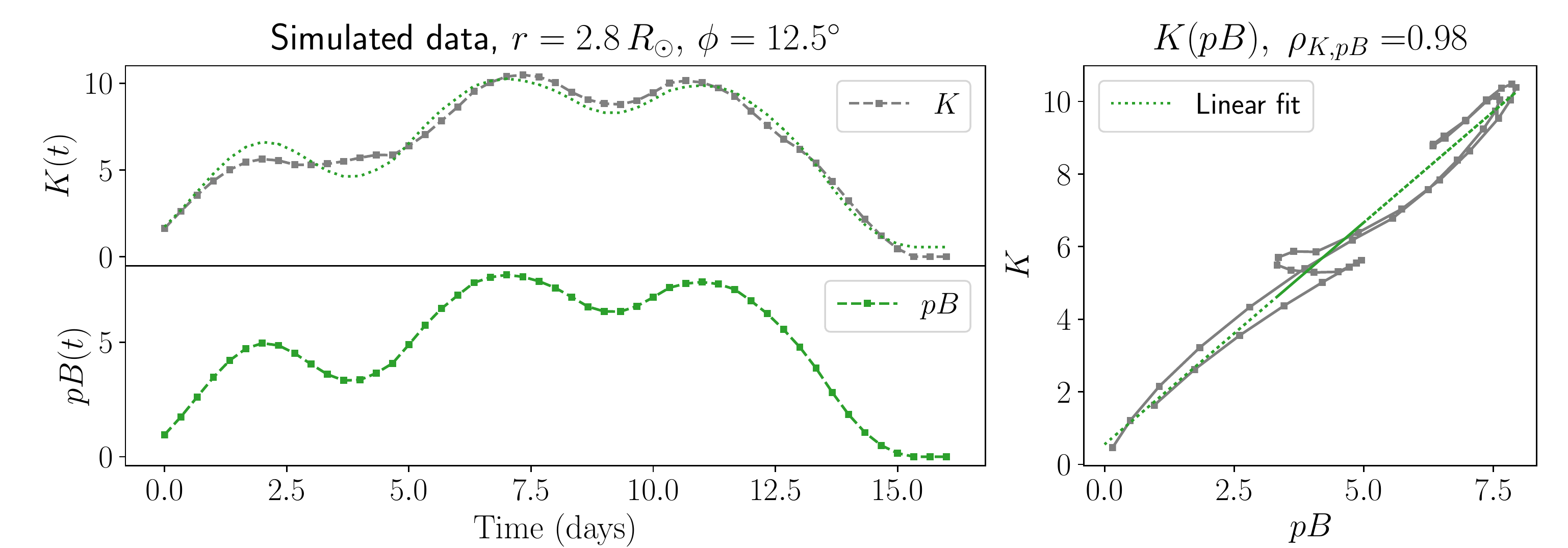} 
        \includegraphics[width=0.43\textwidth]{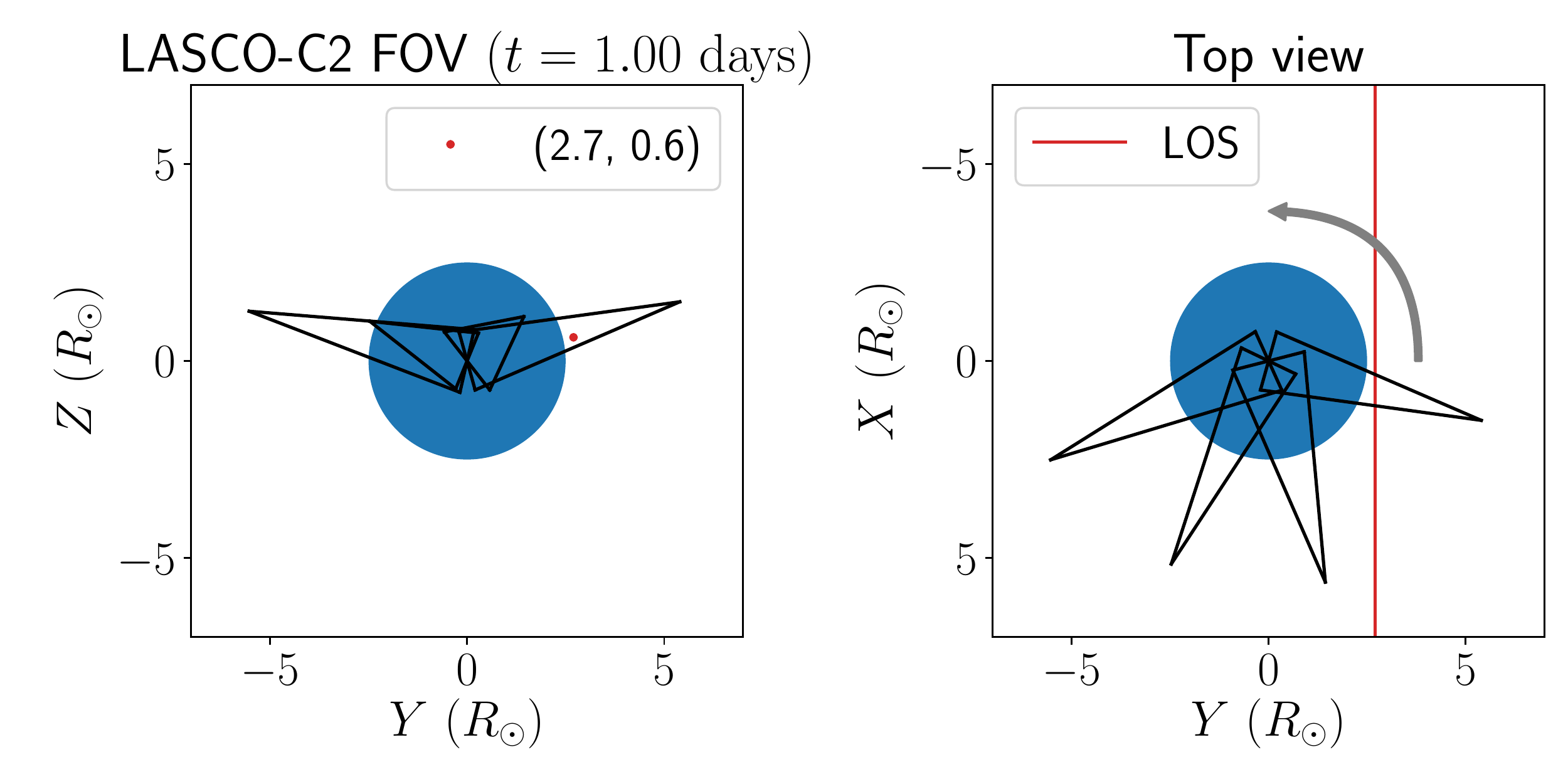} 
        \caption{Same as Fig. \ref{fig:sim_val_int1_broad}, but for the pixel at $r=2.8$ $R_\sun$ and $\phi\approx10\degr$ (March 2008). Bottom panels represent the orientation of four broad simulated streamers at 
        $t=1$ 
        days, as defined in Table \ref{tab:all_sims} (``Sim3'').}
        \label{fig:sim_val_eq_broad} 
\end{figure}

\begin{figure}
        \centering
        \includegraphics[width=0.50\textwidth]{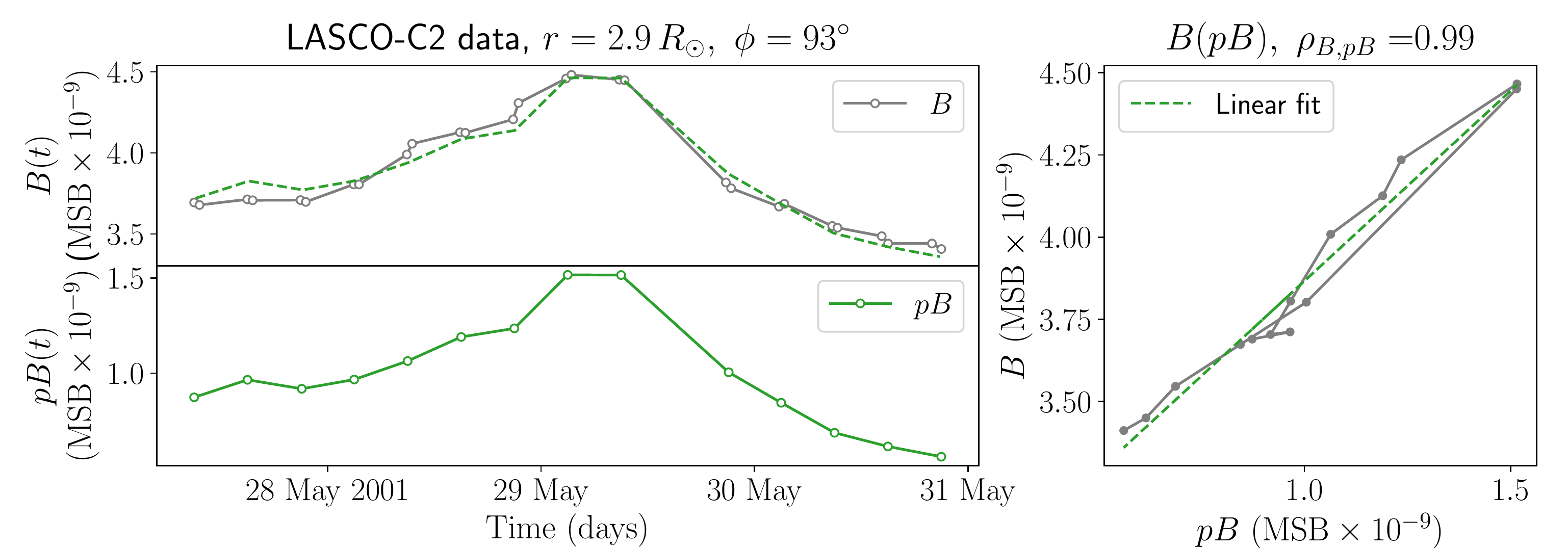} 
        \includegraphics[width=0.50\textwidth]{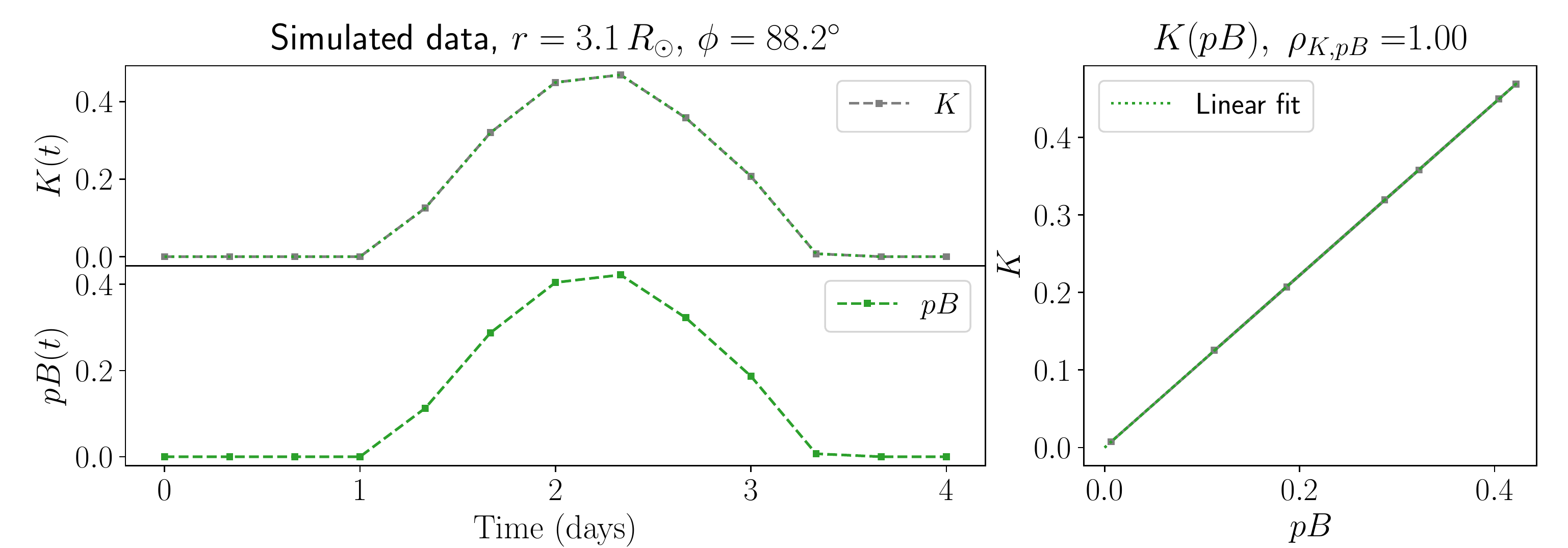} 
        \includegraphics[width=0.43\textwidth]{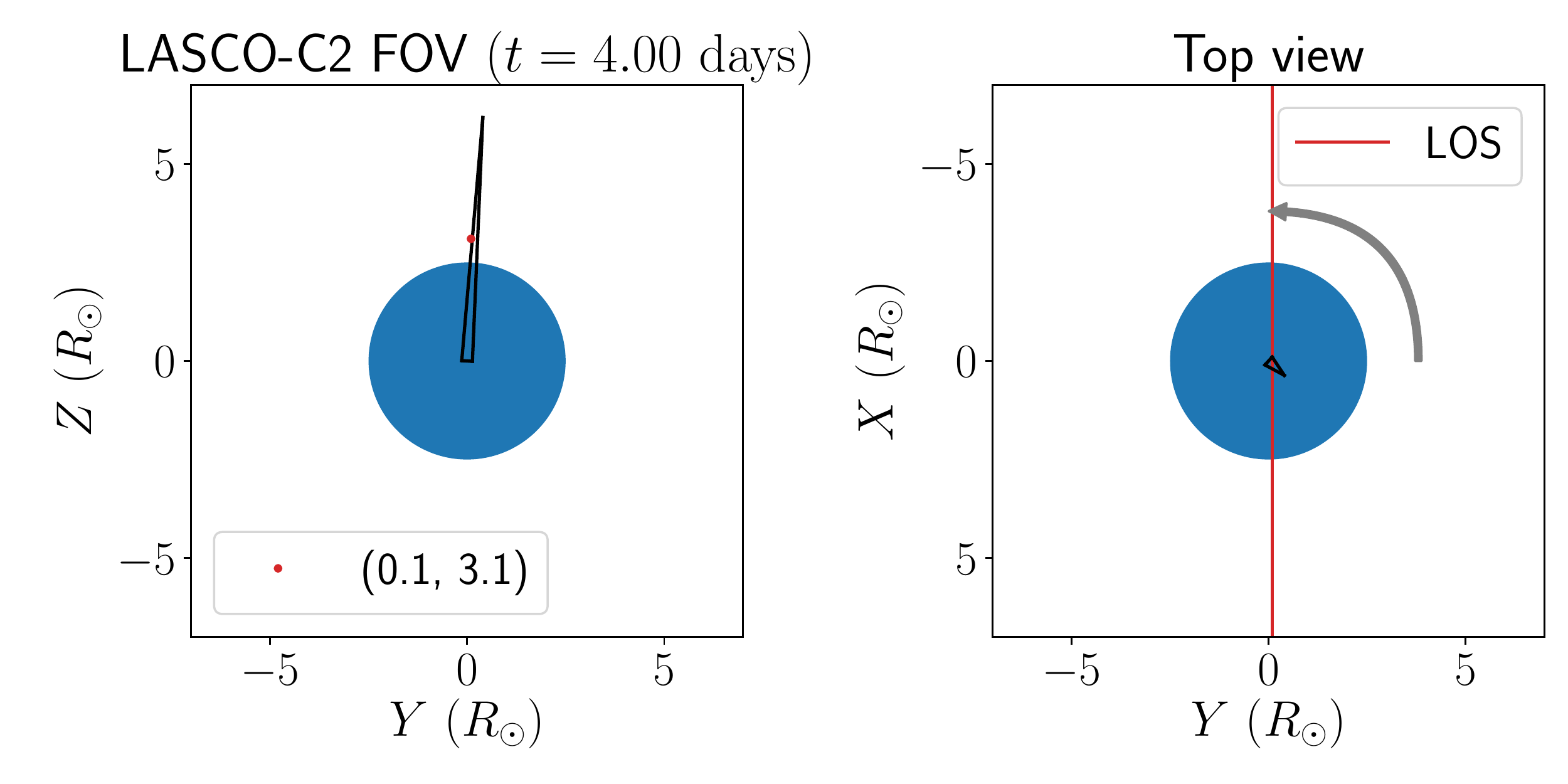} 
        \caption{Same as Fig. \ref{fig:sim_val_int1_broad}, but for the pixel at $r=3.1$ $R_\sun$ and $\phi\approx90\degr$ (May 2001). The real and simulated data cover the time interval of $\sim$4 days. Bottom panels represent the orientation of a narrow simulated streamer at $t=4$ days, as defined in Table \ref{tab:all_sims} (``Sim8'').}
        \label{fig:sim_val_pol_narrow} 
\end{figure}

We simulated the non-evolving streamers that rotate solidly with the Sun and calculated the evolution of $K(t)$ and $pB(t)$ intensities as explained in Sect. \ref{sec:3.2.1}. In our simulations, we considered different numbers of streamers of various sizes and orientations, as well as different pixels within the LASCO-C2 FOV. We compared the evolution of $B$ and $pB$ brightness extracted from the LASCO-C2 data with $K(t)$ and $pB(t)$ simulated for different cases listed in Table \ref{tab:all_sims}. Some examples are shown in Figs. \ref{fig:sim_val_int2_narrow}-\ref{fig:sim_val_pol_narrow}.

\begin{table*}[h]
\caption{Summary of the simulated configurations of streamers.}
\label{tab:all_sims}
\centering
\begin{tabular}{l c c c c c l}
\hline\hline
Label   & POI  [$R_\sun$]       & $\psi(t_0)$   & $\theta$      & $h_{\rm str}$ [$R_\sun$]        & $\beta_{\rm str}$     & Notes \\
\hline
\noalign{\vskip 1mm} 
Sim1            & ($-1.5, 2.4$) & $150\degr$    & $45\degr$     & 6.2   & $20\degr$       & Two broad streamers at intermediate polar angles (Fig. \ref{fig:sim_val_int1_broad}) \\
                        &                               & $135\degr$    & $37.5\degr$     & 5.2   & $18\degr$     &       \\
\noalign{\vskip 1mm} 
Sim2            & ($2.5, 2.5$)  & $-45\degr$    & $52\degr$     & 6.6   & $8.5\degr$& Two narrow streamers at intermediate  polar angles (Fig. \ref{fig:sim_val_int2_narrow})\\
                        &                               & $-75\degr$    & $50\degr$       & 6.2   & $10\degr$     &  \\
\noalign{\vskip 1mm} 
Sim3            & ($2.7, 0.6$)  & $-80\degr$    & $78.3\degr$   & 6.2   & $15\degr$& Four broad equatorial        streamers (Fig. \ref{fig:sim_val_eq_broad}) \\
                        &                               & $-40\degr$    & $80\degr$               & 5.8   & $15\degr$     &    \\
                        &                               & $0\degr$              & $79\degr$               & 5.9   & $18\degr$     &       \\
                        &                               & $60\degr$             & $75\degr$               & 5.8   & $15\degr$     &       \\

\noalign{\vskip 1mm} 
\noalign{\vskip 1mm} 
Sim4            & ($-1.5, 2.4$) & $150\degr$    & $44.4\degr$   & 6.2   & $20\degr$       & Broad streamer at intermediate polar angles \\
\noalign{\vskip 1mm} 
Sim5            & ($-1.5, 2.4$) & $150\degr$    & $37.5\degr$   & 6.2   & $18\degr$       & Broad streamer at intermediate polar angles\\
\noalign{\vskip 1mm} 
Sim6            & ($2.5, 2.5$)  & $-45\degr$    & $52\degr$     & 6.6   & $8.5\degr$      & Narrow streamer at intermediate polar angles\\
\noalign{\vskip 1mm} 
Sim7            & ($2.5, 2.5$)  & $-75\degr$    & $50\degr$     & 6.2   & $10\degr$       & Narrow streamer at intermediate polar angles  \\
\noalign{\vskip 1mm} 
Sim8            & ($0.1, 3.1$)  & $-10\degr$    & $5\degr$      & 6.2   & $2.5\degr$      & Narrow polar streamer (Fig. \ref{fig:sim_val_pol_narrow}) \\
\noalign{\vskip 1mm} 
Sim9            & ($0.1, 3.1$)  & $90\degr$     & $15\degr$     & 6     & $10\degr$       & Narrow polar streamer \\
\noalign{\vskip 1mm} 
Sim10           & ($2.7, 0.6$)  & $-80\degr$    & $78.3\degr$   & 6.2   & $15\degr$       & Broad equatorial streamer     \\
\noalign{\vskip 1mm} 
Sim11           & ($2.7, 0.6$)  & $-30\degr$    & $80\degr$     & 5.8   & $15\degr$       & Broad equatorial streamer     \\
\noalign{\vskip 1mm} 
Sim12           & ($2.7, 0.6$)  & $60\degr$     & $80\degr$     & 5.8   & $15\degr$       & Broad equatorial streamer     \\
\noalign{\vskip 1mm} 
Sim13           & ($2.7, 0.6$)  & $20\degr$     & $79\degr$     & 5.9   & $20\degr$       & Broad equatorial streamer     \\
\hline
\end{tabular}
\tablefoot{Simulations are labeled as shown in the first column. Pixel of interest (POI) is listed in the second column in Cartesian coordinates with respect to the center of the Sun. The third and fourth columns define the orientation of each streamer: $\psi(t_0)$ is the polar angle at the beginning of a simulation (at time $t_0$) and $\theta$ is the azimuthal angle in the spherical coordinate system placed at the center of the Sun ($O$) with the $OX$-axis perpendicular to the POS and pointed toward the observer, and the $OZ$-axis pointing the north pole. The fifth and sixth columns define the size of each streamer: $h_{\rm str}$ is the height and $\beta_{\rm str}$ is the vertex aperture of a streamer. Notes are shown in the last column.}
\end{table*}

\end{appendix}


\begin{thebibliography}{}
\expandafter\ifx\csname natexlab\endcsname\relax\def\natexlab#1{#1}\fi

\bibitem[{{Antonucci} {et~al.}(2005){Antonucci}, {Abbo}, \&
  {Dodero}}]{Antonucci2005}
{Antonucci}, E., {Abbo}, L., \& {Dodero}, M.~A. 2005, \aap, 435, 699

\bibitem[{{Antonucci} {et~al.}(2020){Antonucci}, {Romoli}, {Andretta},
  {Fineschi}, {Heinzel}, {Moses}, {Naletto}, {Nicolini}, {Spadaro}, {Teriaca},
  {Berlicki}, {Capobianco}, {Crescenzio}, {Da Deppo}, {Focardi}, {Frassetto},
  {Heerlein}, {Landini}, {Magli}, {Marco Malvezzi}, {Massone}, {Melich},
  {Nicolosi}, {Noci}, {Pancrazzi}, {Pelizzo}, {Poletto}, {Sasso},
  {Sch{\"u}hle}, {Solanki}, {Strachan}, {Susino}, {Tondello}, {Uslenghi},
  {Woch}, {Abbo}, {Bemporad}, {Casti}, {Dolei}, {Grimani}, {Messerotti},
  {Ricci}, {Straus}, {Telloni}, {Zuppella}, {Auch{\`e}re}, {Bruno},
  {Ciaravella}, {Corso}, {Alvarez Copano}, {Aznar Cuadrado}, {D'Amicis},
  {Enge}, {Gravina}, {Jej{\v{c}}i{\v{c}}}, {Lamy}, {Lanzafame}, {Meierdierks},
  {Papagiannaki}, {Peter}, {Fernandez Rico}, {Giday Sertsu}, {Staub},
  {Tsinganos}, {Velli}, {Ventura}, {Verroi}, {Vial}, {Vives}, {Volpicelli},
  {Werner}, {Zerr}, {Negri}, {Castronuovo}, {Gabrielli}, {Bertacin},
  {Carpentiero}, {Natalucci}, {Marliani}, {Cesa}, {Laget}, {Morea},
  {Pieraccini}, {Radaelli}, {Sandri}, {Sarra}, {Cesare}, {Del Forno}, {Massa},
  {Montabone}, {Mottini}, {Quattropani}, {Schillaci}, {Boccardo}, {Brando},
  {Pandi}, {Baietto}, {Bertone}, {Alvarez-Herrero}, {Garc{\'\i}a Parejo},
  {Cebollero}, {Amoruso}, \& {Centonze}}]{Antonucci2020_metis}
{Antonucci}, E., {Romoli}, M., {Andretta}, V., {et~al.} 2020, \aap, 642, A10

\bibitem[{{Astropy Collaboration} {et~al.}(2018){Astropy Collaboration},
  {Price-Whelan}, {Sip{\H{o}}cz}, {G{\"u}nther}, {Lim}, {Crawford}, {Conseil},
  {Shupe}, {Craig}, {Dencheva}, {Ginsburg}, {VanderPlas}, {Bradley},
  {P{\'e}rez-Su{\'a}rez}, {de Val-Borro}, {Aldcroft}, {Cruz}, {Robitaille},
  {Tollerud}, {Ardelean}, {Babej}, {Bach}, {Bachetti}, {Bakanov}, {Bamford},
  {Barentsen}, {Barmby}, {Baumbach}, {Berry}, {Biscani}, {Boquien}, {Bostroem},
  {Bouma}, {Brammer}, {Bray}, {Breytenbach}, {Buddelmeijer}, {Burke},
  {Calderone}, {Cano Rodr{\'\i}guez}, {Cara}, {Cardoso}, {Cheedella}, {Copin},
  {Corrales}, {Crichton}, {D'Avella}, {Deil}, {Depagne}, {Dietrich}, {Donath},
  {Droettboom}, {Earl}, {Erben}, {Fabbro}, {Ferreira}, {Finethy}, {Fox},
  {Garrison}, {Gibbons}, {Goldstein}, {Gommers}, {Greco}, {Greenfield},
  {Groener}, {Grollier}, {Hagen}, {Hirst}, {Homeier}, {Horton}, {Hosseinzadeh},
  {Hu}, {Hunkeler}, {Ivezi{\'c}}, {Jain}, {Jenness}, {Kanarek}, {Kendrew},
  {Kern}, {Kerzendorf}, {Khvalko}, {King}, {Kirkby}, {Kulkarni}, {Kumar},
  {Lee}, {Lenz}, {Littlefair}, {Ma}, {Macleod}, {Mastropietro}, {McCully},
  {Montagnac}, {Morris}, {Mueller}, {Mumford}, {Muna}, {Murphy}, {Nelson},
  {Nguyen}, {Ninan}, {N{\"o}the}, {Ogaz}, {Oh}, {Parejko}, {Parley}, {Pascual},
  {Patil}, {Patil}, {Plunkett}, {Prochaska}, {Rastogi}, {Reddy Janga},
  {Sabater}, {Sakurikar}, {Seifert}, {Sherbert}, {Sherwood-Taylor}, {Shih},
  {Sick}, {Silbiger}, {Singanamalla}, {Singer}, {Sladen}, {Sooley},
  {Sornarajah}, {Streicher}, {Teuben}, {Thomas}, {Tremblay}, {Turner},
  {Terr{\'o}n}, {van Kerkwijk}, {de la Vega}, {Watkins}, {Weaver}, {Whitmore},
  {Woillez}, {Zabalza}, \& {Astropy Contributors}}]{Astropy2018}
{Astropy Collaboration}, {Price-Whelan}, A.~M., {Sip{\H{o}}cz}, B.~M., {et~al.}
  2018, \aj, 156, 123

\bibitem[{{Astropy Collaboration} {et~al.}(2013){Astropy Collaboration},
  {Robitaille}, {Tollerud}, {Greenfield}, {Droettboom}, {Bray}, {Aldcroft},
  {Davis}, {Ginsburg}, {Price-Whelan}, {Kerzendorf}, {Conley}, {Crighton},
  {Barbary}, {Muna}, {Ferguson}, {Grollier}, {Parikh}, {Nair}, {Unther},
  {Deil}, {Woillez}, {Conseil}, {Kramer}, {Turner}, {Singer}, {Fox}, {Weaver},
  {Zabalza}, {Edwards}, {Azalee Bostroem}, {Burke}, {Casey}, {Crawford},
  {Dencheva}, {Ely}, {Jenness}, {Labrie}, {Lim}, {Pierfederici}, {Pontzen},
  {Ptak}, {Refsdal}, {Servillat}, \& {Streicher}}]{Astropy2013}
{Astropy Collaboration}, {Robitaille}, T.~P., {Tollerud}, E.~J., {et~al.} 2013,
  \aap, 558, A33

\bibitem[{{Blackwell} \& {Petford}(1966)}]{Blackwell1966}
{Blackwell}, D.~E. \& {Petford}, A.~D. 1966, \mnras, 131, 399

\bibitem[{{Boe} {et~al.}(2021){Boe}, {Habbal}, {Downs}, \&
  {Druckm{\"u}ller}}]{Boe2021}
{Boe}, B., {Habbal}, S., {Downs}, C., \& {Druckm{\"u}ller}, M. 2021, \apj, 912,
  44

\bibitem[{{Brueckner} {et~al.}(1995){Brueckner}, {Howard}, {Koomen},
  {Korendyke}, {Michels}, {Moses}, {Socker}, {Dere}, {Lamy}, {Llebaria},
  {Bout}, {Schwenn}, {Simnett}, {Bedford}, \& {Eyles}}]{Brueckner1995}
{Brueckner}, G.~E., {Howard}, R.~A., {Koomen}, M.~J., {et~al.} 1995, \solphys,
  162, 357

\bibitem[{{de Patoul} {et~al.}(2015){de Patoul}, {Foullon}, \&
  {Riley}}]{dePatoul2015}
{de Patoul}, J., {Foullon}, C., \& {Riley}, P. 2015, \apj, 814, 68

\bibitem[{{Dolei} {et~al.}(2015){Dolei}, {Spadaro}, \& {Ventura}}]{Dolei2015}
{Dolei}, S., {Spadaro}, D., \& {Ventura}, R. 2015, \aap, 577, A34

\bibitem[{{Domingo} {et~al.}(1995){Domingo}, {Fleck}, \&
  {Poland}}]{Domingo1995}
{Domingo}, V., {Fleck}, B., \& {Poland}, A.~I. 1995, \solphys, 162, 1

\bibitem[{{Freeland} \& {Handy}(1998)}]{Freeland1998_SSW}
{Freeland}, S.~L. \& {Handy}, B.~N. 1998, \solphys, 182, 497

\bibitem[{{Garc{\'\i}a Marirrodriga} {et~al.}(2021){Garc{\'\i}a Marirrodriga},
  {Pacros}, {Strandmoe}, {Arcioni}, {Arts}, {Ashcroft}, {Ayache}, {Bonnefous},
  {Brahimi}, {Cipriani}, {Damasio}, {De Jong}, {D{\'e}prez}, {Fahmy}, {Fels},
  {Fiebrich}, {Hass}, {Hern{\'a}ndez}, {Icardi}, {Junge}, {Kletzkine}, {Laget},
  {Le Deuff}, {Liebold}, {Lodiot}, {Marliani}, {Mascarello}, {M{\"u}ller},
  {Oganessian}, {Olivier}, {Palombo}, {Philippe}, {Ragnit}, {Ramachandran},
  {S{\'a}nchez P{\'e}rez}, {Stienstra}, {Th{\"u}rey}, {Urwin}, {Wirth}, \&
  {Zouganelis}}]{GarciaMarirrodriga2021_SolarOrbiter}
{Garc{\'\i}a Marirrodriga}, C., {Pacros}, A., {Strandmoe}, S., {et~al.} 2021,
  \aap, 646, A121

\bibitem[{{Gibson} {et~al.}(1999){Gibson}, {Fludra}, {Bagenal}, {Biesecker},
  {del Zanna}, \& {Bromage}}]{Gibson1999}
{Gibson}, S.~E., {Fludra}, A., {Bagenal}, F., {et~al.} 1999, \jgr, 104, 9691

\bibitem[{Harris {et~al.}(2020)Harris, Millman, van~der Walt, Gommers,
  Virtanen, Cournapeau, Wieser, Taylor, Berg, Smith, Kern, Picus, Hoyer, van
  Kerkwijk, Brett, Haldane, Fernández~del Río, Wiebe, Peterson,
  Gérard-Marchant, Sheppard, Reddy, Weckesser, Abbasi, Gohlke, \&
  Oliphant}]{2020NumPy-Array}
Harris, C.~R., Millman, K.~J., van~der Walt, S.~J., {et~al.} 2020, Nature, 585,
  357

\bibitem[{{Hayes} {et~al.}(2001){Hayes}, {Vourlidas}, \& {Howard}}]{Hayes2001}
{Hayes}, A.~P., {Vourlidas}, A., \& {Howard}, R.~A. 2001, \apj, 548, 1081

\bibitem[{{Howard} {et~al.}(2019){Howard}, {Vourlidas}, {Bothmer}, {Colaninno},
  {DeForest}, {Gallagher}, {Hall}, {Hess}, {Higginson}, {Korendyke},
  {Kouloumvakos}, {Lamy}, {Liewer}, {Linker}, {Linton}, {Penteado}, {Plunkett},
  {Poirier}, {Raouafi}, {Rich}, {Rochus}, {Rouillard}, {Socker}, {Stenborg},
  {Thernisien}, \& {Viall}}]{Howard2019}
{Howard}, R.~A., {Vourlidas}, A., {Bothmer}, V., {et~al.} 2019, \nat, 576, 232

\bibitem[{Hunter(2007)}]{Matplotlib2007}
Hunter, J.~D. 2007, Computing In Science \& Engineering, 9, 90

\bibitem[{{Koutchmy} \& {Lamy}(1985)}]{KoutchmyLamy1985}
{Koutchmy}, S. \& {Lamy}, P.~L. 1985, {The F-Corona and the Circum-Solar Dust
  Evidences and Properties (ir)}, ed. R.~H. {Giese} \& P.~{Lamy}, 63

\bibitem[{{Lamy} {et~al.}(2021){Lamy}, {Gilardy}, {Llebaria}, {Qu{\'e}merais},
  \& {Ernandez}}]{Lamy2021_C3_K_F}
{Lamy}, P., {Gilardy}, H., {Llebaria}, A., {Qu{\'e}merais}, E., \& {Ernandez},
  F. 2021, \solphys, 296, 76

\bibitem[{{Lamy} {et~al.}(2020){Lamy}, {Llebaria}, {Boclet}, {Gilardy},
  {Burtin}, \& {Floyd}}]{Lamy2020_C2}
{Lamy}, P., {Llebaria}, A., {Boclet}, B., {et~al.} 2020, \solphys, 295, 89

\bibitem[{{Lamy} {et~al.}(1997){Lamy}, {Quemerais}, {Llebaria}, {Bout},
  {Howard}, {Schwenn}, \& {Simnett}}]{Lamy1997}
{Lamy}, P., {Quemerais}, E., {Llebaria}, A., {et~al.} 1997, in ESA Special
  Publication, Vol. 404, Fifth SOHO Workshop: The Corona and Solar Wind Near
  Minimum Activity, ed. A.~{Wilson}, 491

\bibitem[{{Llebaria} {et~al.}(2021){Llebaria}, {Lamy}, {Gilardy}, {Boclet}, \&
  {Loirat}}]{Llebaria2021_F_SL}
{Llebaria}, A., {Lamy}, P., {Gilardy}, H., {Boclet}, B., \& {Loirat}, J. 2021,
  \solphys, 296, 53

\bibitem[{{Minnaert}(1930)}]{Minnaert1930}
{Minnaert}, M. 1930, \zap, 1, 209

\bibitem[{{Morgan} \& {Habbal}(2010)}]{Morgan2010}
{Morgan}, H. \& {Habbal}, S. 2010, \apj, 711, 631

\bibitem[{{Morgan} \& {Habbal}(2007)}]{Morgan2007}
{Morgan}, H. \& {Habbal}, S.~R. 2007, \aap, 471, L47

\bibitem[{{Morrill} {et~al.}(2006){Morrill}, {Korendyke}, {Brueckner},
  {Giovane}, {Howard}, {Koomen}, {Moses}, {Plunkett}, {Vourlidas},
  {Esfandiari}, {Rich}, {Wang}, {Thernisien}, {Lamy}, {Llebaria}, {Biesecker},
  {Michels}, {Gong}, \& {Andrews}}]{Morrill2006}
{Morrill}, J.~S., {Korendyke}, C.~M., {Brueckner}, G.~E., {et~al.} 2006,
  \solphys, 233, 331

\bibitem[{{Qu{\'e}merais} \& {Lamy}(2002)}]{Quemerais2002}
{Qu{\'e}merais}, E. \& {Lamy}, P. 2002, \aap, 393, 295

\bibitem[{{Ragot} \& {Kahler}(2003)}]{Ragot2003}
{Ragot}, B.~R. \& {Kahler}, S.~W. 2003, \apj, 594, 1049

\bibitem[{{Saito} {et~al.}(1977){Saito}, {Poland}, \& {Munro}}]{Saito1977}
{Saito}, K., {Poland}, A.~I., \& {Munro}, R.~H. 1977, \solphys, 55, 121

\bibitem[{{Stauffer} {et~al.}(2018){Stauffer}, {Stenborg}, \&
  {Howard}}]{Stauffer2018}
{Stauffer}, J.~R., {Stenborg}, G., \& {Howard}, R.~A. 2018, \apj, 864, 29

\bibitem[{{Stenborg} \& {Howard}(2017)}]{Stenborg2017}
{Stenborg}, G. \& {Howard}, R.~A. 2017, \apj, 839, 68

\bibitem[{{van de Hulst}(1950)}]{vandeHulst1950}
{van de Hulst}, H.~C. 1950, \bain, 11, 135

\bibitem[{Virtanen {et~al.}(2020)Virtanen, Gommers, Oliphant, Haberland, Reddy,
  Cournapeau, Burovski, Peterson, Weckesser, Bright, {van der Walt}, Brett,
  Wilson, Millman, Mayorov, Nelson, Jones, Kern, Larson, Carey, Polat, Feng,
  Moore, {VanderPlas}, Laxalde, Perktold, Cimrman, Henriksen, Quintero, Harris,
  Archibald, Ribeiro, Pedregosa, {van Mulbregt}, \& {SciPy 1.0
  Contributors}}]{2020SciPy-NMeth}
Virtanen, P., Gommers, R., Oliphant, T.~E., {et~al.} 2020, Nature Methods, 17,
  261

\end{thebibliography}
\end{document}